\documentclass[pra,twocolumn,nofootinbib,superscriptaddress,showpacs,floatfix]{revtex4-1}
\usepackage{amsmath,amssymb,amsfonts}
\usepackage{mathcomp}
\usepackage{comment}
\usepackage{graphicx,subfigure}
\usepackage{txfonts}
\usepackage{epstopdf}
\usepackage[title]{appendix}
\usepackage{float}

\makeatletter
\newsavebox{\@brx}
\newcommand{\llangle}[1][]{\savebox{\@brx}{\(\m@th{#1\langle}\)}%
  \mathopen{\copy\@brx\kern-0.5\wd\@brx\usebox{\@brx}}}
\newcommand{\rrangle}[1][]{\savebox{\@brx}{\(\m@th{#1\rangle}\)}%
  \mathclose{\copy\@brx\kern-0.5\wd\@brx\usebox{\@brx}}}
\makeatother

\usepackage[usenames, dvipsnames]{color}
\usepackage{soul}
\usepackage[normalem]{ulem}
\usepackage{physics}
\usepackage{bbold}
\usepackage{footnote}

\usepackage{xr-hyper} 
\usepackage[unicode]{hyperref} 
\hypersetup{
	unicode=true,                                           
	plainpages=false,
	pdffitwindow=true,                                      
	colorlinks=true,                                        
	linkcolor=blue,                                        
	citecolor=blue,                                         
	filecolor=blue,                                        
	urlcolor=blue                                          
}

\newcommand*\colvec[3][]{
	
	\begin{pmatrix}\ifx\relax#1\relax\else#1\\\fi#2\\#3\end{pmatrix}
}

\def\bbraket#1{\mathinner{\llangle{#1}\rrangle}}
\def\kket#1{\mathinner{\vert{#1}\rrangle}}

\begin{document}
       \title{Two-colour photon correlations probe coherent vibronic contributions to electronic excitation transport under incoherent illumination} 
        \author{Charlie Nation}
        \email{c.nation@ucl.ac.uk}
        \affiliation{Department of Physics and Astronomy, University College London, Gower Street, WC1E 6BT, London, United Kingdom}
 	\author{Valentina Notararigo}
        \affiliation{Department of Physics and Astronomy, University College London, Gower Street, WC1E 6BT, London, United Kingdom}
        \author{Hallmann Óskar Gestsson}
        \affiliation{Department of Physics and Astronomy, University College London, Gower Street, WC1E 6BT, London, United Kingdom}
        \author{Luca Sapienza}
        \affiliation{Department of Engineering, University of Cambridge, Cambridge, CB3 0FA, United Kingdom}
	\author{Alexandra Olaya-Castro}
	\email{a.olaya@ucl.ac.uk}
	\affiliation{Department of Physics and Astronomy, University College London, Gower Street, WC1E 6BT, London, United Kingdom}
	
	\date {\today}

 	\begin{abstract}
Identifying signatures of quantum coherent behaviour in photoactive systems that are maintained in stationary states away from thermal equilibrium is an open problem of wide interest in a variety of physical scenarios, including single photosynthetic complexes subjected to continuous incoherent illumination. Here we consider a prototype light-harvesting heterodimer exhibiting coherent and collective exciton-vibration interactions and show that the second-order frequency-filtered correlations of fluorescence photons provide insightful information on the influence of such coherent interactions for different transitions, thereby yielding fundamentally different photon-counting statistics. Furthermore, we show that coherent vibronic mechanisms strongly affect the asymmetries characteristic of time-resolved photon cross-correlations and manifest themselves in a time-dependent violation of the Cauchy-Schwarz inequality bounding cross-correlations for classically fluctuating fields. We finally discuss how such second-order correlation asymmetry establishes important connections between coherent vibronic interactions, directional exciton population transport, and violation of quantum detailed balance. Our work then indicates that measurement of two-colour photon correlation asymmetry can be an important avenue to investigate quantum behaviour of single photoactive biomolecular and chemical systems under incoherent illumination conditions. 

	\end{abstract}

	\maketitle
\section{Introduction}

Open quantum systems maintained in stationary states away from thermal equilibrium are of fundamental importance in a variety of fields ranging from quantum optics \cite{Carmichael2, landi2023} and quantum thermodynamics \cite{vinjanampathy2016, deffner2019} to quantum processes in biomolecular and chemical systems \cite{fassioli2012, huelga2013, romero2014, romero2017, scholes2017, kim2021}. Central in each case is the determination of when the open system of interest may exploit quantum coherent interactions in a non-equilibrium steady state (NESS), and the signatures of such quantum behaviour in an accessible experimental settings. Indeed, systems in a NESS are crucial for investigating the implications of quantum coherent mechanisms in violating classical limits such as the violation of Cauchy-Schwartz inequality under a semi-classical description of the correlations of emitted light \cite{clauser1974, loudon1980, Reid1986, Peiris2015} or the violation of quantum detailed balance (QDB), which characterises microscopic reversibility of dynamics \cite{denisov2002, tomita1973}.

Photosynthetic light-harvesting complexes absorb sunlight and efficiently direct the associated collective electronic excitation (exciton) energy towards a reaction centre where chemical energy conversion takes place \cite{Amerongen2000, Blankenship2002}.  This excitation transport process relies on a combination of coherent and incoherent interactions between electronic and vibrational motions  \cite{kolli2012,fassioli2012, chin2012, huelga2013, Oreilly2014}. Of particular interest in the last decade has been the influence of vibrational motions whose energy scale commensurate to energy differences between excitonic transitions  \cite{Oreilly2014, higgins2021, higgins2021a, calderon2023, nation2023}. Such coherent vibronic interactions have been shown to potentially aid exciton transport in a quantum coherent manner while exhibiting non-classical features \cite{Oreilly2014}. The question yet remains as to reliable probes of such quantum behaviour in single photosynthetic biomolecules when they are subjected to incoherent illumination and de-excitation processes. Notably, characterisation of NESS photoexcitation transport under incoherent illumination enables understanding of these photoactive biomolecules close to their natural operational conditions.  

It is well known that photon counting theory and experiments are key to identify quantum behaviour of the light emitted and of the emitting source \cite{Glauber, grangier_1986, glauber2006, Lounis2000, olaya-castro2001, Michler2000, Moreau2001}. In fact, recent works have suggested that photon correlations \cite{iles-smith2017, cygorek2023, wiercinski2023} and cross-correlations \cite{Holdaway2018, SanchezMunoz2020, cygorek2023, downing2023, humphries2023} may witness cooperative and coherent behaviour of complex quantum emitters. Experiments have also demonstrated anti-bunching in single molecule room-temperature measurements of bichromophoric \cite{hofkens2003, hubner2003, kim2019} and multi-chromophoric \cite{Wientjes2014} systems. Photon-correlations measurements have also very recently been exploited to demonstrate single-photon activation processes in photosynthetic complexes \cite{li2023}. However, how photon correlations may specifically witness the coherent exciton-vibration mechanisms underpinning excitation transport in biomolecular systems, and more generally, the interplay between coherent and incoherent processes, has not yet been investigated.

In this work we consider a prototype light-harvesting heterodimer and investigate how markers of coherent exciton-vibration interactions underpinning excitation transport processes under incoherent illumination may be probed via measurements of \textit{frequency-filtered} photon-correlations (FFPCs). We focus on the second-order FFPCs \cite{Eberly1977, DelValle2012, Holdaway2018, phillips2020, Gonzalez-Tudela2013a, Peiris2015, ngaha} of the emitted light in the steady-state, both at zero-time and at a finite-time delay $\tau$.  We show that introducing a frequency filter at around specific optical transitions allows the probing of different types of collective mechanisms, which lead to fundamentally different photon-counting statistics as function of the exciton-vibration coupling. Time-resolved second-order FFPCs exhibit a time-asymmetry when the two filters have differing frequencies. We demonstrate that such correlation asymmetry is strongly sensitive to coherent interactions between excitons and vibrations and that the sequence in which photo detection takes place uncover time-dependent violations of the Cauchy-Schwartz inequality, which are significantly stronger in the presence of such coherent vibronic interactions. We further interpret such correlation asymmetry in terms of violations of quantum detail balance, thereby establishing connections between coherent vibronic interactions, directional excitation transport, and non-classical behaviour under incoherent illumination. Our results therefore put forward two-color correlation time-asymmetry as promising approach to investigate quantum processes and the underpinning electronic-vibrational interactions in a variety of single photoactive biomolecules and complex chemical systems under NESS. 

\section{Results}

\subsection{Vibronic dimer model}\label{sec:model}

In this work we consider a biologically-inspired vibronic heterodimer \cite{Tiwari2012, Oreilly2014, Killoran2015, Dean2016, nation2023, calderon2023, chuang2023}, in which each chromophore (site) has an excited electronic state $|k\rangle$, with energy $\epsilon_k$, $k=1,2$, with inter-site coupling strength $V$. Each site is locally coupled with strength $g_v$ to a quantised vibrational mode of frequency $\omega_\textrm{vib}$. The vibronic Hamiltonian is then $H=H_{\textrm{el}}+ H_{\textrm{vib}}+H_{\textrm{el-vib}}$, where
	$ H_{el} = \epsilon_1 |1\rangle \langle 1| + \epsilon_2 |2\rangle \langle 2| + (\epsilon_1  + \epsilon_2)|1, 2\rangle \langle 1, 2| + V(|1\rangle \langle 2| + |2\rangle \langle 1|)$ with $|1, 2\rangle \langle 1, 2|$ the double excited state where sites 1 and 2 are simultaneously excited. 
	$H_{\textrm{vib}}= ~\omega_{\textrm{vib}}(d_1^\dag d_1+ ~d_2^\dag d_2)$,
	and
	$ H_{\textrm{el-vib}} ~=~ g_v \sum_{k=1,2}|k\rangle \langle k| (d^\dag_k+ d_k)$.
	Here $d^\dag_k$($d_k$) creates (annihilates) a phonon of the vibrational mode of site $k$. 
 
\begin{figure*}
    \includegraphics[width=0.98\textwidth]{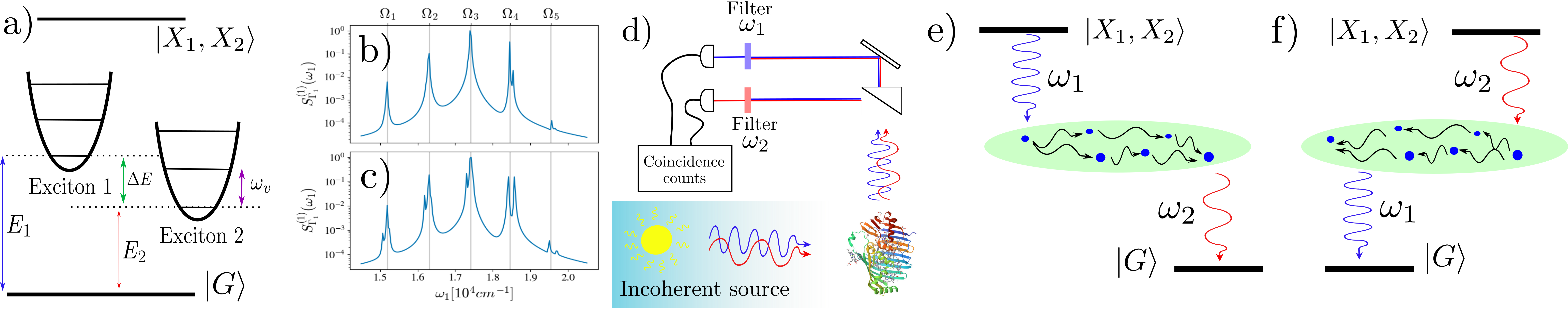}
    \caption{Diagram of setup and characterisation of prototype photosynthetic dimer model. Note vibrational levels of ground and doubly excited states not shown. a) Diagram of electronic energy levels of dimer model. b), c) Show physical spectrum \cite{Eberly1977} of dimer 1 and 2 respectively. d) Shows a diagram of experimental proposal. e), f) Show positive and negative time cross correlation contributions, respectively. }
    \label{fig:diagram}
\end{figure*}

We define the excitonic eigenstates of $H_{el}$ as $\{|G\rangle, |X_1\rangle, |X_2\rangle, |X_1, X_2\rangle \}$, 
 where $|G\rangle$ is the state where both chromophores are in the ground state, $|X_1\rangle$ and $|X_2\rangle$ are the excitonic eigenstates of  
	$H_{\textrm{el}}$ in the single excitation subspace, with corresponding eigenenergies $E_1$ and $E_2$ (with $E_1 > E_2$, see Fig. \ref{fig:diagram}a)) and an energy gap 
	$\Delta E =\sqrt{(\Delta \epsilon)^2+4V^2}$, with $ \Delta \epsilon=\epsilon_1-\epsilon_2$
	being the positive difference between the onsite energies. $|X_1, X_2\rangle$ is the double exciton eigenstate with eigenenergy  $E_1+E_2$. Transformation of $H$ into collective exciton states and normal mode coordinates \cite{Oreilly2014}, with creation operators
	$D_{\textrm{com}/\textrm{rd}}^{(\dag)}=(d^{(\dag)}_1 \pm d^{(\dag)}_2)/\sqrt{2}$,
	leads to the Hamiltonian taking the form of a generalised quantum Rabi model \cite{Xie2017}:
	\begin{equation}\begin{split}\label{H_diag}
	H_0 = & \, E \, M + \frac{\Delta E}{2} \: \sigma_z + 
	\omega_{\textrm{vib}} (D^\dag_{\textrm{com}} D_{\textrm{com}} + D^\dag_{\textrm{rd}} D_{\textrm{rd}})\\ 
	& \; + \frac{g_v}{\sqrt{2}} \: 
	\Big(\cos(2\theta) \, \sigma_z - \sin(2\theta) \, \sigma_x\Big) (D_{\textrm{rd}}+D_{\textrm{rd}}^\dag)\\ 
	& \; + \frac{g_{v}}{\sqrt{2}} M  \:  (D_{\textrm{com}}+D_{\textrm{com}}^\dag).
	\end{split}\end{equation}
	Here we have defined the collective electronic operators 
	$M~=~ |X_1\rangle \langle X_1| + |X_2\rangle\langle X_2| + 2|X_1, X_2\rangle \langle X_1, X_2|$,
	$\sigma_z= |X_1\rangle \langle X_1| - |X_2\rangle\langle X_2|$, 
	and 
	$\sigma_x =|X_2\rangle \langle X_1| +|X_1\rangle \langle X_2| $, 
	and the mixing angle $\theta~=~1/2 \arctan \left(2|V|/\Delta\epsilon \right)$ satisfies $0<\theta<\pi/4$. The center of mass (COM) mode $D_{\textrm{com}}$ can be seen to thus cause only an energy shift of the excitonic states, but is not involved directly in excitonic transitions. The relative displacement (RD) mode $D_{rd}$, however, is coupled directly to excitonic transitions, and thus directly contributes to exciton transport. 
 
 The collective vibrational eigenstates of $D^\dag D = D^\dag_{\textrm{com}}D_{\textrm{com}} + D^\dag_{\textrm{rd}}D_{\textrm{rd}}$ are labelled $|l\rangle$ and, for numerical computation their Hilbert space is truncated to a maximum number $2L$, i.e. $l=0,1,\cdots,(L_{rd} + L_{com})$, with a maximum occupation $L_{rd}(L_{com})$ in the RD(COM) mode. Convergence is expected for a reasonably small $L$, as $\omega_v \gg k_B T \sim g_v$, such that the (maximum considered) vibrational coupling and temperature are not large enough to highly populate the vibrational modes. Indeed, we observe that $L_{rd} = 6$, $L_{com} = 4$ is enough to ensure convergence in most cases, which we use unless explicitly stated. Ground vibronic eigenstates of $H_0$ are of the form $|G, l\rangle\equiv|G\rangle\otimes |l\rangle$ while an excited vibronic eigenstate of energy $v$ are labelled $|F_v\rangle$ and can be written as quantum superpositions of states $|X_k,l\rangle\equiv|X_k\rangle\otimes| l\rangle$ i.e. $|F_v\rangle~=~\sum_{l=0}^{2L}\sum_{k=1,2}C_{kl}(v)|X_k, l \rangle$.
                    
	Since we have considered the direct influence of the most relevant vibrational motion at the Hamiltonian level, the dissipative dynamics induced by the surrounding environment as well as incoherent excitation and emission are treated via Lindblad superoperators \cite{BP_c3}. We are interested in the case where $g_v\ll\Delta E\sim\omega_{vib}$, as occurs in several light-harvesting proteins \cite{schulze2016, caycedo-soler2018, klinger2020, higgins2021a}. For these conditions the system can be subjected to incoherent pumping of the highest energy exciton $|X_1\rangle$ with jump operator $\sigma_{X_1}^{\dag} = |X_1\rangle \langle G| + |X_1, X_2\rangle \langle X_2|$ and rate $P_{X_1}$. Radiative decay processes occur from excited vibronic states to the ground state at rate $\gamma_{\nu, l} = F_{\nu, l} \gamma$, where $F_{\nu, l} = \sum_m \langle m, l| F_\nu \rangle$ \cite{nation2023}, and with associated jump operators $\sigma_{v,n}=|G,n\rangle \langle F_v|$. Similar dissipative processes from doubly excited to singly excited states occur at rates $F_{v,l}^{(2)} = \sum_m \langle F_\nu | m, l\rangle $ via jump operator $\sigma_{v,n}^{(2)} = |F_v \rangle \langle X_1, X_2, l|$ (see SM \cite{SM} Section I.D for derivation of dissipative processes). We assume each site undergoes pure dephasing at a rate $\gamma_{pd}$ with jump operator $A_k=|k\rangle \langle k| + |k, \overline{k}\rangle \langle k, \overline{k}|$, where $k \in [1, 2]$, with $|\overline{2}\rangle = |1\rangle$ and vice versa. The collective vibrations (RD and COM modes) each undergo thermal emission and absorption processes with rates $\Gamma_{th} (\eta(\omega_{vib})+1)$ and $\Gamma_{th} \eta (\omega_{vib})$, respectively. Here $\eta(\omega)~=~\left(e^{\beta\omega}-1\right )^{-1}$ with $\beta=1/k_B T$ the thermal energy scale. The Liouvillian of the vibronic heterodimer under these incoherent processes is given by
	\begin{align}\label{eq:systemME}
	\mathcal{L}_{0}( &\hat{\rho}) = - i \,[H_0,\hat{\rho}]  \,  + \sum_{k={1,2}} \frac{\gamma_{pd}}{2} \mathcal{L}_{A_k}(\hat{\rho}) \nonumber\\
	&  + \sum_{D \in [D_{rd}, D_{com}]}\left( \frac{\Gamma_{th} (\eta(\omega_{vib})+1)}{2} \, \mathcal{L}_{D}(\hat{\rho}) 
	+\frac{\Gamma_{th} \eta (\omega_{vib})}{2} \, \mathcal{L}_{D^{\dag}}(\hat{\rho})\right) \nonumber \\
	& + \, \frac{\gamma}{2} \, \sum_{v=1}^{2L} \sum_{l=1}^{L} (\,F_{\nu, l}  \mathcal{L}_{\sigma_{vl}}(\hat{\rho}) + F^{(2)}_{\nu, l}  \mathcal{L}_{\sigma_{vl}^{(2)}}(\hat{\rho})  )
	\,+ \, \frac{P_{X_1}}{2} \mathcal{L}_{\sigma^{\dag}_{X_1}}(\hat{\rho}),
	\end{align}
	where  $\mathcal{L}_c(O)~ = ~(2cOc^{\dag} - c^{\dag}cO - Oc^{\dag}c)$. 
	
	We consider this model in two parameter regimes detailed in Table \ref{table:dimer_params}. The first, labelled dimer 1, is chosen to resemble chryptophyte antennae Phycoerythrin 545  \cite{collini2010, curutchet2013}. Dimer 2 is chosen to have identical excitonic energies to dimer 1, whilst displaying increased excitonic delocalisation.  We will vary $g_v$ to investigate the change in the photon correlations as a function of the coherent coupling strength  between excitons and  collective vibrational motions. 
    A diagram of the 
    excitonic states of each dimer is shown in Fig. \ref{fig:diagram}a), and its physical spectrum is presented in Fig. \ref{fig:diagram}b, c). Here we note that the spectra is peaked around (vibrationally dressed) excitonic energies $\Omega_3, \Omega_4$, and smaller peaks occur at intervals of the vibrational frequency. In the following the peaked emission frequencies $\Omega_3$ and $\Omega_4$ will be chosen as the filtered frequencies. Note also that inclusion of the doubly excited state in the above model is vital in order to capture the correct behaviour of second-order photon correlations, which require the emission of two photons.
    
	 \begin{table}[h]
	\begin{tabular}{ |c|c|c| } 
		\hline
		& Dimer 1 & Dimer 2 \\ 
		\hline 
		$\zeta = \frac{2V}{\Delta\alpha}$ & 0.1 & 0.5 \\ 
		$V$ & 92cm${}^{-1}$ & 236.6cm${}^{-1}$ \\ 
		$\Delta\alpha$ & 1042cm${}^{-1}$ & 946.5cm${}^{-1}$ \\ 
		\hline
		\multicolumn{3}{|c|}{Common Parameters}\\
		\hline
		& $\Delta E = 1058.2$cm${}^{-1}$ & $\gamma$ = 1ns${}^{-1} \approx $ 0.01 cm$^{-1}$  \\
		& $E = 18000$cm${}^{-1}$ & $\gamma_{pd} =$ 10ps${}^{-1} \approx$ 5.31 cm$^{-1}$ \\
		& $\omega_{vib} = 1111$cm${}^{-1}$ & $P_{X_1}=$ $5\times 10^{-4}$ ps${}^{-1} \approx$ 0.003 cm$^{-1}$ \\
		& $g_v=267.1$cm${}^{-1}$ & $\Gamma_{th}=$1 ps${}^{-1} \approx$ 5.31 cm$^{-1}$ \\
		& $\beta := (k_BT)^{-1} = 300$K & $\Gamma$ = 0.2 ps${}^{-1} \approx$ 1.06 cm$^{-1}$ \\
		\hline
	\end{tabular}
	\caption{\label{table:dimer_params}Parameters of dimer 1 and dimer 2. Top section shows parameters that differ between the models, bottom shows those that are shared. Where the parameters deviate from those listed here we mention explicitly.}
\end{table}

The above model has two core mechanisms through which exciton transport can occur. The first, which we label the incoherent transfer mechanism, is due to the pure dephasing environmental process. This mechanism has been studied in great detail in e.g. Refs. \cite{caruso2009, chin2010, chin2012}, and is often labelled environmentally assisted quantum transport (ENAQT). Here, excitonic transfer is enabled by pure dephasing acting to localise excitonic excitations to their component sites, due to the intersite coupling $V$ excitonic transfer may then occur, and thus an initial excitation in state $|X_1\rangle$ may transfer to $|X_2\rangle$ in the presence of this incoherent process.  
We note, that whilst we label the above mechanism incoherent, this mechanism nonetheless relies on the formation of delocalised excitons and intersite couplings $V$.

The second mechanism of exciton transport in the above model we label the coherent transfer mechanism, which can be understood as a consequence of a single term in Eq. \eqref{H_diag}:
\begin{align}\label{eq:Hterm}
    \frac{g_v}{\sqrt{2}} \sin (2\theta) \sigma_x (D_{rd} + D^\dagger_{rd}). 
\end{align}
This term implies that displacements of the RD collective vibrational mode enable excitonic transitions through $\sigma_x = |X_1\rangle \langle X_2| + |X_2\rangle \langle X_1|$. We also notice that excitonic delocalisation is necessary for this mechanism to occur, inasmuch as $\theta$ must be sufficient for the effective coupling strength $\frac{g_v}{\sqrt{2}} \sin(2\theta)$ to contribute significantly. The RD mode thus plays the central role in coherent excitonic transfer, and this core mechanism can be captured in excitation dynamics by ignoring the COM mode entirely (see SM \cite{SM}). However the COM mode has a considerable effect on photon emission, as it couples differently to different excitation number subspaces. 

In order to show that, under continuous incoherent pumping of the emitters,  the collective RD mode assisting exciton transport is indeed a non-classical channel, in the SM, Supplemental Figures \ref{fig:coherences} and \ref{fig:coherences_omegas}, we provide results on the steady-state exciton coherences and vibrational number fluctuations
It has previously been shown that in a transient regime, this collective mode exhibits non-classical thermal fluctuations \cite{Oreilly2014}. In Supplemental Figures \ref{fig:coherences}  and \ref{fig:coherences_omegas} we show that this statement also holds true for the NESS studied here. We further confirm in Supplemental Figure \ref{fig:dynamics} that coupling to such vibrational motion indeed leads to an increased exciton population transfer, and and increased excitonic coherence in the transient regime. 

\subsection{Frequency-filtered photon correlations}
FFPCs are defined by a time and normal ordered correlation \cite{cresser1987, knoll1986},
\begin{equation}\label{eq:FFPCdef}
g^{(2)}_{\Gamma_1\Gamma_2}(\omega_1, T_1; \omega_2, T_2) =\frac{S_{\Gamma_1\Gamma_2}^{(2)}(\omega_1, T_1; \omega_2, T_2)} {S_{\Gamma_1}^{(1)}(\omega_1, T_1) S_{\Gamma_1}^{(2)}(\omega_2, T_2)} \;,
\end{equation}
with $S^{(1)}(\omega, T_1) = \frac{\Gamma_1}{2\pi}e^{-\Gamma_1 T_1} \iint_{-\infty}^{T_1} F_{\omega_1, \Gamma_1}(t_1)F^*_{\omega_1, \Gamma_1}(t_2) \langle a^\dagger(t_1)a(t_2)\rangle$, 
defining the physical spectrum \cite{Eberly1977}, and $S^{(2)}(\omega_1, T_1; \omega_2, T_2) \\ = \frac{\Gamma_1\Gamma_2}{4\pi^2}e^{-\Gamma_1 (T_1+T_2)} \iint_{-\infty}^{T_1}dt_1 dt_2\iint_{-\infty}^{T_2}ds_1 ds_2 F_{\omega_1, \Gamma_1}(t_1)F^*_{\omega_1, \Gamma_1}(t_2)\\  \times F_{\omega_2, \Gamma_2}(s_1) F^*_{\omega_2, \Gamma_2}(s_2)\langle \mathcal{T}_-[a^\dagger(t_2) a^\dagger(s_1)]  \mathcal{T}_+[a(t_2) a(s_2)] \rangle$
with $\mathcal{T}_-$ and $\mathcal{T}_+$ the time-ordering and anti-ordering superoperators.  
Here $F_{\omega, \Gamma}(t) = e^{-\frac{\Gamma}{2}t - i\omega t}$ defines Lorentz filters, and $a$ are system operators which couple to the detected light field.
Since we are interested in the steady state regime, the physical spectrum is time independent, and two-photon correlations depend only on the delay $\tau = T_1 - T_2$. We further assume for simplicity identical filter linewidths  i.e. $\Gamma_1 = \Gamma_2$. 

In general, the intensity-intensity cross correlations quantified by $g^{(2)}_{\Gamma}(\omega_1, \omega_2, \tau)$ are time-asymmetric when $\omega_1\neq \omega_2$ i.e. 
$g^{(2)}_{\Gamma}(\omega_1, \omega_2, |\tau|) \neq g^{(2)}_{\Gamma}(\omega_1, \omega_2, -|\tau|)$, as
the photon correlation for negative $\tau$ corresponds to the conditional probability of detecting a photon of frequency $\omega_1$ given that a photon of frequency $\omega_2$ was detected a zero-delay, and for positive $\tau$ the reverse measurement order is implied. In contrast, when  $\omega_1= \omega_2$, $g^{(2)}_{\Gamma}(\omega_1, \omega_1, \tau)$ are time-symmetric.

 FFPCs of the above form have been of both theoretical and experimental interest for many decades. Beginning with Eberly and Wódkiewicz, who generalised the theoretical description of optical spectra to non-stationary systems, or physical spectrum \cite{Eberly1977}, which we show in Figs. \ref{fig:diagram}b) and c) for each dimer. Higher order correlations were formalised in Refs. \cite{cresser1987, knoll1986}, which show that in order to obtain the correct time-ordering consideration of the measurement apparatus itself is necessary. Asymmetry in FFPCs was observed by Aspect et. al \cite{aspect1980}, and later by Schrama et. al \cite{schrama1992}, where explicit calculations described this asymmetry as a result of coherent coupling in the effective Hamiltonian. Theoretical works, however, were limited due to the complexity of evaluating the four-dimensional time ordered integral of Eq. \eqref{eq:FFPCdef}. More recently, the development of a sensor method \cite{DelValle2012, Holdaway2018} has enabled efficient calculations of FFPCs in a wide array of systems.  As described in the Methods section, our computations of FFPCs follow the perturbative formulation of the sensor method which some of us developed in \cite{Holdaway2018}, and in the SM  \cite{SM} we show new efficient algebraic solutions for the underlying integrals presented in \cite{Holdaway2018}.
 
\subsection{Broad-band correlations}\label{sec:broad_band}

\begin{figure}
    \includegraphics[width=0.98\linewidth]{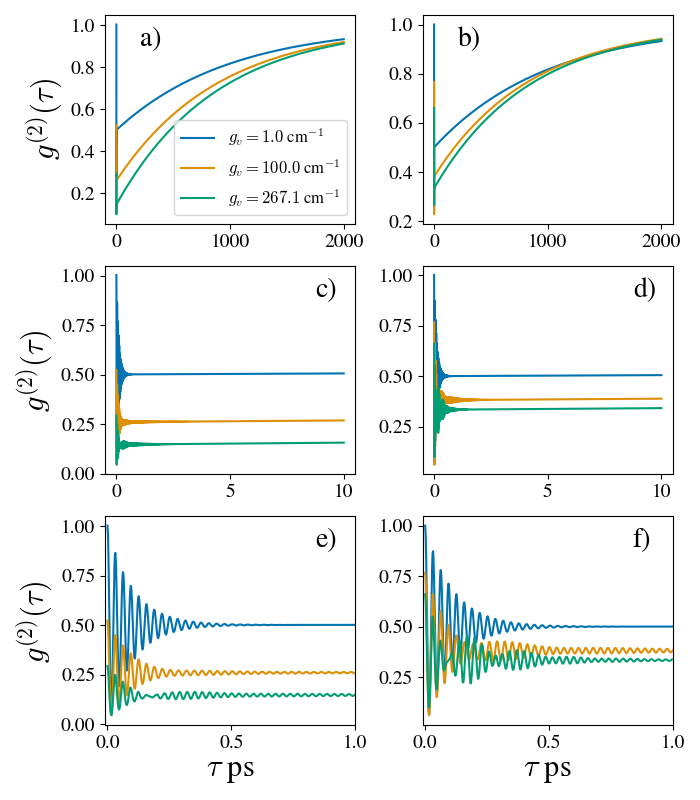}
    \caption{Broad-band photon-counting statistics for a), c), e) dimer 1 and b), d), f) dimer 2. Same data shown at differing timescales. }
    \label{fig:regular_g2}
\end{figure}

Broad-band photon correlations are given by the $\Gamma\to \infty$ limit of Eq. \eqref{eq:FFPCdef}: $g^{(2)}(\tau) = \frac{\langle a^\dagger a^\dagger(\tau) a(\tau) a\rangle}{|\langle a^\dagger a \rangle|^2}$, and are shown for the two heterodimers under considerations in Fig. \ref{fig:regular_g2}. We observe initial transient oscillations decaying over the pure-dephasing timescale  (Figs. \ref{fig:regular_g2}e)) and \ref{fig:regular_g2}f)), followed by a quasi-stationary state reached over $\sim 1$ picosecond (Figs. \ref{fig:regular_g2}c) and \ref{fig:regular_g2}d)). The quasi-stationary value is due to equilibration in the single excitation  subspace,$\{|X_1\rangle, |X_2\rangle \}$, after the first emission generates a non-equilibrium state i.e. $\rho_{eff} \propto a \rho a^\dagger$. After this quasi-equilibration timescale, broad band correlations only purvey information on the excited state life time over much longer,  nanosecond timescales (Figs. \ref{fig:regular_g2}a) and \ref{fig:regular_g2}b)).  

As the fast oscillations $g^{(2)}(\tau)$ occur over far too short a timescale to be experimentally measurable (see Section \ref{sec:experiment} for further discussion),  we can state that the key information in the short-time scales can be inferred from the quasi-equilibrium value for $g^{(2)}(\tau_{eq})$ (Fig. 2c and 2d). Both, $g^{(2)}(0)$ (observed at the peak of the first oscillation) and $g^{(2)}(\tau_{eq})$, depend sensitively on the vibrational coupling strength, with increased coupling $g_v$ acting to \emph{antibunch} photon emission. Complete antibunching, $g^{(2)}(0) = 0$, is the definitive hallmark of the system behaving as a two-level emitter. Hence, these results indicate that  coherent exciton-vibration interaction increases the directional exciton transfer to the lower energy exciton state in the steady-state, leading the system to act closer to a two-level system as $g_v$ is increased.

We also observe a dependence of broad-band correlations on the degree of exciton delocalisation and therefore their cooperative behavior.  Dimer 2 has a slightly larger exciton delocalisation than dimer 1, while retaining  in both cases a quasi-localised exciton structure.  We observe very little change in the values for  $g^{(2)}(0)$ when $g_v = 1$ cm$^{-1}$, with $g^{(2)}(0) \approx 1.0030$ for dimer 1 and $g^{(2)}(0) \approx 1.0003$ for dimer 2, and in each case having quasi-equilibrium values  $g^{(2)}(\tau_{eq})$ $\sim 0.5$, which is the value expected for two dephased uncoupled two-level systems \cite{auffeves2011}. For larger $g_v$, we see that dimer 1 is the more antibunched (both in the zero-delay value and in the quasi-stationary value). We thus observe that $g^{(2)}(0)$ and  $g^{(2)}(\tau_{eq})$ demonstrate a competition between excitonic cooperativity leading to bunching in photoemission \cite{loudon1980, auffeves2011, SanchezMunoz2020} and coherent exciton-vibration coupling that selectively populates the lowest exciton state and leads to antibunched emission. 

Interestingly, previous experimental work has measured imperfect antibunching in zero-delay broad-band photon correlations on  the light emitted by a large photosynthetic complex  isolated from purple bacteria (LH2) which exhibits two distinct absorption bands weakly coupled \cite{Wientjes2014}. Our results here suggest that such experimental result may indeed be demonstrating the competing mechanisms here discussed, and that accounting accurately for the influence of phonon environments will be crucial in rationalising such experimental result.

Similar phenomena have been studied for cooperative emission of identical two-level systems in a cavity \cite{auffeves2011}, where cooperative and individualised regimes were observed via zero-delay $g^{(2)}$. In our case, the competing effects of vibronic coupling and excitonic cooperativity may not be easy to discern in an experimental measurement of broadband $g^{(2)}(0)$. In the next section we discuss  how spectral filtering and FFPCs may provide additional information of such interplay by probing different transfer pathways. 

\subsection{Zero-delay frequency filtered photon correlations}\label{sec:g20FFPC}

By frequency filtering the light emission, even zero-delay correlations contain some information on time dependence. This is due to the definition in Eq. \eqref{eq:FFPCdef}, as the zero-delay value requires time integrations over a timescale associated with the filter linewidth $\Gamma$. Thus, increased frequency resolution (smaller $\Gamma$) of an emitted photon reduces the possible certainty in its time of detection.
We see an example in Figs. \ref{fig:g20}a), b), where we scan $g^{(2)}(\Omega_3, \omega_1, 0)$ over the emission frequencies $\omega_1$ for each dimer. We observe that correlations peak for values corresponding to vibrational and excitonic resonances of the system in a similar manner to the physical spectrum in Fig. \ref{fig:diagram}b). We observe an overall shift towards lower values of $g^{(2)}(\Omega_3, \omega_1, 0)$ as $g_v$ is increased, aside from the peaks where bunching is observed in some cases. This corroborates the previous observation that the quasi-coherent vibrational  motion supports a behaviour close to that of a two-level quantum emitter. Notably, we observe that these FFPCs are particularly sensitive to exciton-vibration interaction. 

\begin{figure}
    \includegraphics[width=0.98\linewidth]{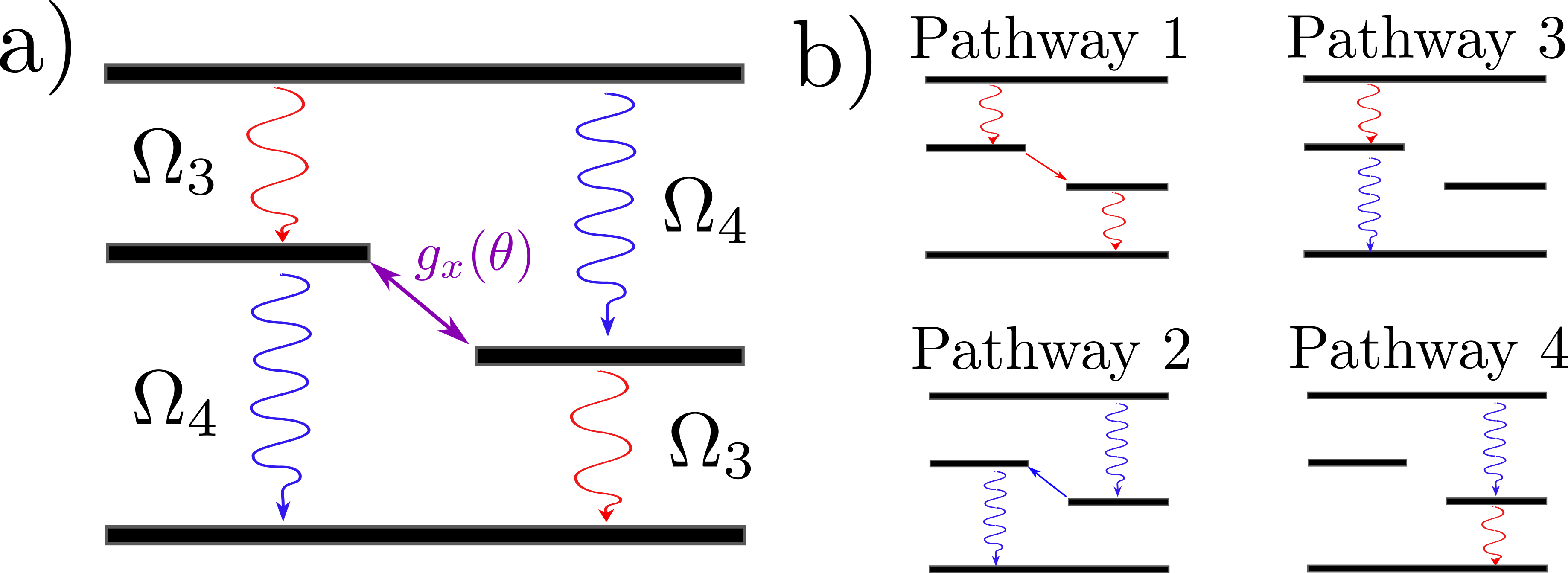}
    \caption{Dominant transitions leading to auto- and cross-correlations in the dimer model around excitonic energies a) shows each transition with labelled frequency and effect coherent transfer $g_x(\theta) = \frac{g_v}{2}\sin(\theta)$. b) Shows each available pathway for two photon emission at short times (where re-pumping can be ignored). Pathways 1 and 2 correspond to autocorrelations, and pathways 3 and 4 correspond to cross-correlations.}
    \label{fig:g20_pathways}
\end{figure}

\begin{figure}
    \includegraphics[width=0.98\linewidth]{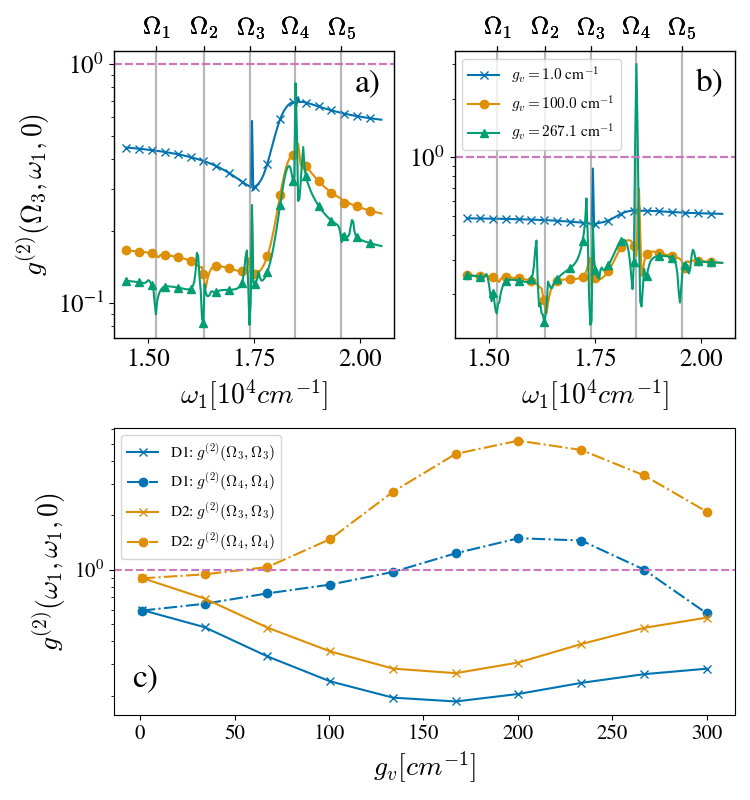}
    \caption{Zero delay FFPCs for FFPCs. a), b) show a scan of the filter frequency $\omega_2$, conditioned on a detection of $\omega = \Omega_3$, which is the excitation energy of the highest energy exciton.  dimer 1 and dimer 2, respectively. Red dotted horizontal line shows coherence bound from Ref. \cite{SanchezMunoz2020} of $g^{(2)} \leq 1$, magenta dashed line shows uncorrelated value $g^{(2)} = 1$. c) Shows autocorrelations for photon pairs of energies ($\Omega_3, \Omega_3$) and ($\Omega_4, \Omega_4$) for dimer 1 (D1) and dimer 2 (D2).}
    \label{fig:g20}
\end{figure}

Such sensitivity to vibronic coupling is further analysed in Fig. \ref{fig:g20}c), where we show how particular autocorrelations, chosen at emission peaks $\Omega_3, \Omega_4$ near exciton transitions, are altered by $g_v$. We illustrate the pathways corresponding to autocorrelations $g^{(2)}(\Omega_3, \Omega_3, 0)$ and $g^{(2)}(\Omega_4, \Omega_4, 0)$ in Fig. \ref{fig:g20_pathways}b) (pathways 1 and 2, respectively). For increasing $g_v$ the behaviour of zero-delay auto-correlations differs between the two filters; at $g_v = 0$ the values are near identical, whereas larger $g_v$ values lead to bunching from the higher energy emission, and antibunching of the lower energy emission with a non-monotonic behaviour. 

\begin{figure*}
    \includegraphics[width=0.98\linewidth]{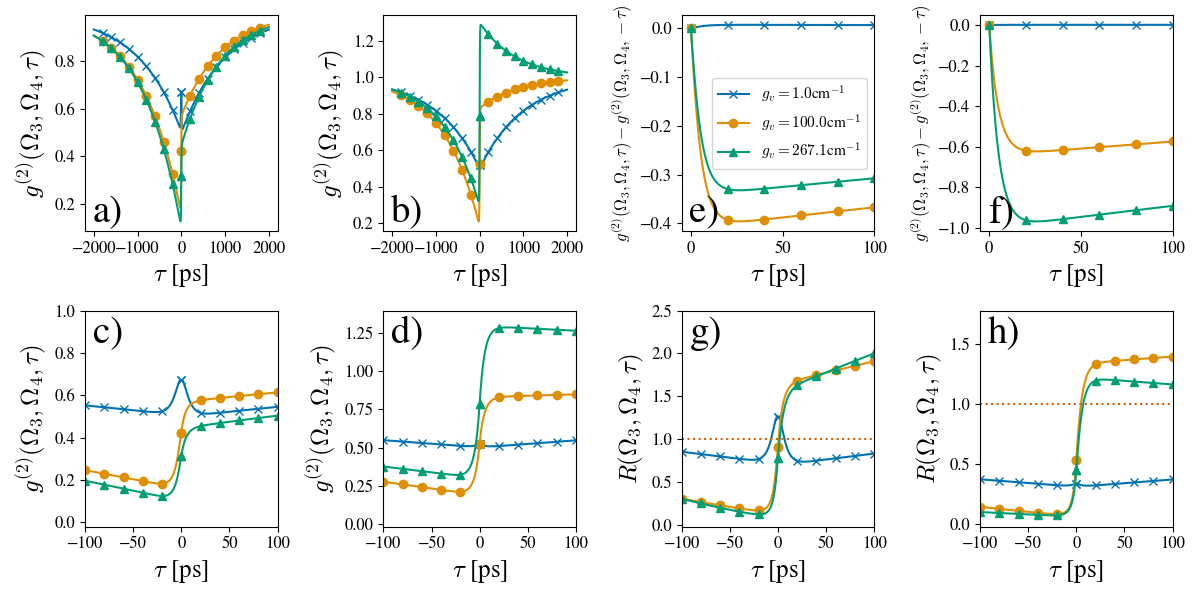}
    \caption{Time dependence of FFPCs (a), b)), and their early time evolution (c), d)). PPFC asymmetry shown in (e), f)), and the CSI measure Eq. \eqref{eq:CSI_measure} (g), h)) for dimer 1 (b), f) d), h)) and dimer 2 (a), c) e), g)).}
    \label{fig:g2t}
\end{figure*}

We thus observe that the two filters probe different phenomena present in the model. The low energy filter around $\Omega_3$, probing Pathway 1 in Fig. \ref{fig:g20_pathways}b), shows a tendency to anti-bunching as $g_v$ increases. This is due to directional vibration-assisted population transfer  from $|X_1\rangle$ to $|X_2\rangle$. Contrarily, in the high energy filter we probe Pathway 2 in Fig. \ref{fig:g20_pathways}b), which due to lower relative occupation of $|X_1\rangle$ displays bunching of emitted light, as the contribution of the doubly excited state to emission of light at frequency $\Omega_4$ is increased. These are the same mechanisms that we see to affect broad-band correlations above  which are able to discern via different frequency filters.

Comparing dimers 1 and 2, we see that relatively small increase in exciton delocalisation of dimer 2 results in a significant deviation induced by $g_v$ between the two autocorrelations,  due to larger collective vibronic coupling  strength $\frac{g_v}{\sqrt{2}} \sin(2\theta)$  dominating the coherent transfer mechanism (Eq. \eqref{eq:Hterm}). We observe that each pathway is additionally bunched (or less anti-bunched) for dimer 2 compared to dimer 1 as $g_v$ increases, essentially indicating that each pathway still witnesses cooperative behaviour associated to large exciton delocalisation.  In Supplemental Figure \ref{fig:g20_gpd} of the SM \cite{SM} we show the dependence of the selected auto-correlations in Fig. \ref{fig:g20}c) on $\gamma_{pd}$ with and without the coherent vibrational mode. We observe that the autocorrelations only substantially differ if coherent vibrational coupling is present, and that increasing $\gamma_{pd}$ acts only to make autocorrelations coalesce. Thus, crucially, the observed deviation of exciton auto-correlations in Fig. \ref{fig:g20}c) is seen to be a direct consequence of the coherent transport mechanism.

We have thus seen that the zero-delay auto-correlations can give insightful information on the influence of coherent vibronic couplings on different transfer pathways, yet they alone are not enough to certify the non-classical nature of the dynamical processes underpinning the excitation transport, for which time-resolved measurements of FFPCs will be needed \cite{Reid1986}. We thus turn to discussing how additional information on quantum transport is accessible via analysis of the time dependence of such correlations. 

\subsection{Correlation Asymmetry}\label{sec:asymmetry}

The central unique aspect of photon cross-correlations is that their time dependence may be asymmetric on swapping the order of observations. In Figs. \ref{fig:g2t}a) -d) we present $g^{(2)}(\Omega_3, \Omega_4, \tau)$ for the two dimers of interest and different $g_v$ values, and observe an overall increase in correlation asymmetry as $g_v$ is increased. This asymmetry is further highlighted in Figs. \ref{fig:g2t}e) and f) where we plot the difference  $g^{(2)}(\Omega_3, \Omega_4, \tau)-g^{(2)}(\Omega_3, \Omega_4, -\tau)$ for the short-time scale of few hundred of picoseconds. While such difference is larger as $g_v$ increases, it is a non-monotonic function of $g_v$. This is due to competing effects of the two collective vibrational modes coupled via $g_v$ (see Eq. \eqref{H_diag}). The RD mode contributes directly to coherent exciton transport as described above. The COM mode couples in far simpler manner but, while it does not contribute to exciton transport \cite{Tiwari2012}, it does affect photoemission and witnesses coherent vibronic coupling. 
The COM mode has therefore a more subtle influence, which may be understood as a renormalisation of the energy of the exciton states \cite{nation2023}. In Supplemental Figure \ref{fig:no_com} of the SM \cite{SM} we present the time-resolved FFPCs discarding the effect of the COM mode, and confirming that in such a situation increasing $g_v$  results in a monotonic increase in FFPC asymmetry as $g_v$ increases. We will analyse the effect of the COM mode further in the next section.

FFPCs may additionally act to probe non-classicality via a violation of the Cauchy-Schwartz inequality (CSI) for cross-correlations of classically radiating fields \cite{loudon1980, Reid1986, SanchezMunoz2014a}, that is, by violating the upper limit $R \leq 1$, with 
\begin{align}\label{eq:CSI_measure}
    R(\omega_1, \omega_2, \tau) = \frac{g^{(2)}(\omega_1, \omega_2, \tau)^2}{g^{(2)}(\omega_1, \omega_1, 0) g^{(2)}(\omega_2, \omega_2, 0)}.
\end{align}
In Fig. \ref{fig:g2t}g), h) we show $R(\omega_3, \omega_4, \tau)$ vs $\tau$ for dimers 1 and 2, respectively. For each dimer $R(\omega_3, \omega_4, \tau)$ is itself asymmetric in time with $R(\omega_3, \omega_4, \tau>1)$ for $\tau>0$ for finite $g_v$, and yields additional asymmetry as $g_v$ is increased (which we note is again monotonic upon removal of the COM mode as shown in the SM). Notice that for $g_v\sim 0$ violation of the classical inequality in Eq. \eqref{eq:CSI_measure} only happens at zero-delay for dimer 1 where excitons are quasilocalised. We thus observe directionality in the non-classicality of photon cross-correlations is induced by coherent vibronic coupling. 

In order to demonstrate that the observed asymmetries are indeed a consequence of the coherent transfer mechanism, in Supplemental figure \ref{fig:g2t_gpd1} we study the effect of pure dephasing on correlation asymmetry to assess the impact of the incoherent transfer mechanism.  We do not observe a significant change to the asymmetry with $\gamma_{pd}$ for small $g_v$, and the observed asymmetries as $g_v$ is increased follow similar behaviour to Fig. \ref{fig:g2t}. This indicates that the observed asymmetry is dominated by the coherent transfer mechanism, meaning that photon cross-correlation asymmetry can indeed distinguish coherent and incoherent transfer processes. 

It is key to highlight that whilst correlation asymmetry here is dominated by the coherent transfer process, and that the asymmetry lasts for the entire nanosecond timescale over which excitations in the system decay, this does not imply that in a transient situation coherences in the system will last for this long timescale. Indeed, we show in the SM, Fig. \ref{fig:dynamics}c), f) the transient excitation dynamics illustrating that coherences are completely decayed over $\sim 1 $ ps. 

Crucially, what these results indicate is that coherent transfer acts to set up a correlation asymmetry in the short timescale (see Fig. \ref{fig:g2t}c), d)) and that such correlations then slowly tend to the uncorrelated value over nanosecond timescales. Thus, the consequence of early time coherent transfer is observed for these long timescales, where coherences themselves have long decayed. Importantly, the FFPC asymmetry is extremely sensitive to collective vibronic coupling, thereby witnessing the transport directionality provided by this mechanism. We will shortly discuss how then this asymmetry can be more formally linked to arguments of microreversibility and QDB violation. Overall, these results then show how insightful FFPC asymmetry measurements can be.

\subsection{Exploration of off-resonant conditions}\label{sec:off_resonant}

In the analysis so far we have assumed that energy scales of the exciton gaps and the vibrational frequency commensurate i.e. $\omega_v \sim \Delta E$. However, both experimental \cite{Dean2016, higgins2021, higgins2021a} and theoretical \cite{kolli2012, Oreilly2014,Caycedo-Soler2022, calderon2023} works have highlighted the influence of a variety of vibrational motions in the ultrafast photoexcitation dynamics in the systems of interest. Here we investigate the effect of relaxing such vibrational resonance condition i.e.  $\omega_v \neq \Delta E$ on the FFPC asymmetry. 

\begin{figure}
    \includegraphics[width=0.48\textwidth]{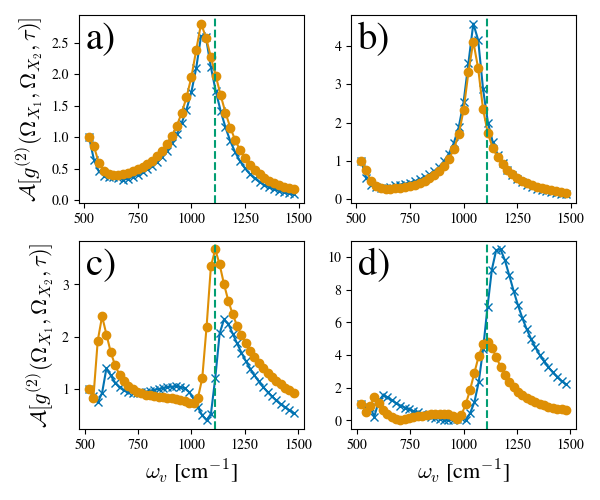}
    \caption{Total asymmetry $\mathcal{A}[g^{(2)}(\omega_1, \omega_2, \tau), \tau]$ of photon correlations out to $\tau =$ 10 ps for a), c) dimer 1 and b), d) dimer 2. We show a constant coupling $g_v = 100$ cm$^{-1}$ in a), b) and $g_v = 267.1$ cm$^{-1}$ in c), d). Here we compare the asymmetry in the presence (blue crosses) and absence (orange circles) of the COM mode. Vertical dashed green line show at 1111 cm$^{-1}$, the vibrational frequency in Fig. \ref{fig:g2t}.}
    \label{fig:g2t_omegas}
\end{figure}

We define the total asymmetry $
    \mathcal{A}[f(t)] = ||f(t) - f(-t)||,
   $
   where $|| f(t) || = \sqrt{h\sum_k f(t_k)}$, $h = \Delta t / N_{points}$, with $\Delta t$ and $N_{points}$ the range in and number of time values, respectively. In Fig. \ref{fig:g2t_omegas} we show $\mathcal{A}[f(t)]$ as function of the vibrational frequency $\omega_v$ while keeping all other parameters fixed. 
   For the medium coupling strength $g_v = 100$ cm$^{-1}$ shown in Fig. \ref{fig:g2t_omegas}a), b) for dimer 1 and dimer 2, respectively, we see that there is a resonance peak in the asymmetry when the vibrational frequency is close to that of the exciton energy gap. We additionally observe a smaller peak in the asymmetry around 530 cm$^{-1}$, which we associate to a resonance of the harmonic. There is no significant difference caused by exclusion of the COM mode in this case. 
   
   For the stronger coupling strength of $g_v = 267.1$ cm$^{-1}$ we observe more complex behaviour when comparing the asymmetry in the presence and absence of the COM, as shown in Fig. \ref{fig:g2t_omegas}c), d).  The COM mode causes a notable shift in the vibrational frequency that leads to maximal asymmetry. Further, the presence of this mode in dimer 1 reduces the resonance effect overall, while in dimer 2 in-fact increases the maximal asymmetry. 
   The source of the non-monotonic behaviour of asymmetry observed in Fig. \ref{fig:g2t} for dimer 1 can thus be traced to the significant shift in the resonance peak frequency with $g_v$, and thus intermediate values of $g_v$ may lie closer to resonance for the chosen $\omega_v$.

   We can thus conclude that resonance conditions between exciton energy gap and specific vibrational frequencies do favour asymmetry of the two-color FFPCs for the heterodimer under consideration, witnessing an associated to directional excitation transport. In the next section we discussed how these observations can then be directly linked to violation of microreversibility conditions.
   
\subsection{Connecting photon correlations to microreversibility, transport, and quantum detailed balance}\label{sec:QDB}

\begin{figure}
    \includegraphics[width=0.48\textwidth]{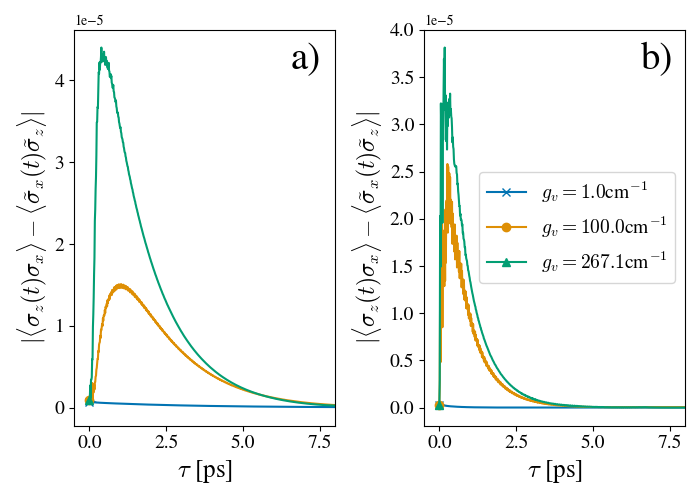} 
    \caption{Violation of QDB as measured by mixed population-coherence correlations for a) dimer 1 and b) dimer 2. Non-zero values indicate a violation of QDB in the NESS.}
    \label{fig:qdb}
\end{figure}

We have seen above that the asymmetry of photon cross-correlations is able to witness directional excitation transfer induced by a coherent and collective vibronic transfer mechanism. We show in Methods \ref{sec:microrev_g2} that such correlation asymmetry is related to the \emph{microreversibility} of dynamics.

Microreversibility is the property of a system with time-reversal symmetry of dynamics \cite{carmichael1976}. Unitary evolution is trivially reversible in this sense, and open systems can violate microreversibility, indicated via a violation of QDB. There are multiple definitions of QDB, however here we follow Refs. \cite{agarwal1973, carmichael1976}, using what was labelled in Ref. \cite{roberts2021} as `conventional' QDB, which is satisfied if we have the equality:
 \begin{equation}\label{eq:qdb_def}
    \langle A(t)B(0)\rangle \stackrel{QDB}{=} \langle \tilde{B}^\dagger(t) \tilde{A}^\dagger(0)\rangle,
\end{equation}
for all observables $A, B$, where the expectation values are taken in the steady-state $\rho_{ss}$ of the underlying dynamical generator $\mathcal{L}$, such that $\dot{\rho}_{ss} = \mathcal{L}[\rho_{ss}] = 0$, and $\tilde{A} = TAT^{-1}$ is the time-reversed form of the operator $A$, where $T$ is the anti-unitary time reversal operator. We show in the Methods that the same microreversibility conditions leading to symmetric photon correlations in the case, as studied here, of reversible observables $\tilde{A} = A$ (which follows from the real symmetric form of the Hamiltonian), lead to satisfaction of an equality of the form \eqref{eq:qdb_def} in a doubled Hilbert space (see Methods \ref{sec:choi}). For completeness we provide a discussion of the time reversal operator $T$ in the SM.

In the remainder of this section we will study the impact of coherent vibrational modes on the violation of QDB as measured via Eq. \eqref{eq:qdb_def}. 
In order to probe the coherent contribution to QDB violation we choose $A = \sigma_x = |X_1\rangle \langle X_2| + |X_2\rangle \langle X_1|$, and $B = \sigma_z = |X_1\rangle \langle X_1| - |X_2 \rangle \langle X_2|$. Here the $\sigma_x$ observable directly probes excitonic coherences. Crucially, in Fig. \ref{fig:qdb} we observe that, indeed, the coherent transfer mechanism induces an additional channel through which QDB is violated. We note that this is unaffected by the COM mode due to the choice of observables, and hence the violation monotonically increases with $g_v$. 

We note that increasing $V$ for some choices of observables $A, B$ in fact \emph{decreases} QDB violations, due to an increased unitary contribution to overall dynamics, and hence a greater degree of reversibility in the dynamics. This increase in unitarity occurs also for dimer 2 compared with dimer 1, as the increase in the Hamiltonian coupling parameters relative to incoherent environmental influence leads to increased reversibility of dynamics. This effect is show in Supplemental Figure \ref{fig:qdb_pops} for QDB measured via population operators only. This shows that the chosen observables which manifest a violation of QDB may each probe differing mechanisms in the system which do so. We observe in Fig. \ref{fig:qdb} that $g_v$ increases QDB violation as the vibronic coupling enables an additional mechanism of coherent transfer through which QDB is violated. 

\subsection{Experimental feasibility}\label{sec:experiment}

Measurement of the FFPC asymmetries shown in Fig. \ref{fig:g2t} requires a suitable time resolution on the order of picoseconds. We note that correlation measurements have been carried out successfully using streak cameras, that can reach time resolutions less than 800 fs \cite{wiersig2009, assmann2009}, or superconducting nanowire single-photon detectors, whose time resolutions can be down to a few picoseconds \cite{korzh2020, esmaeilzadeh2020}. Both these detection systems would prove sufficient for the experimental verification of the effects predicted in this work. More pronounced asymmetries, like the ones shown, for instance, in Fig. \ref{fig:g2t}, could be observed even using lower-cost, commercial avalanche photodiodes that can have time resolutions of the order of 35 ps.

Concerning the experimental evaluation of the role of vibrational modes in the photoexcitation transport and in the consequent variation in the FFPC asymmetry, we note that phononic crystals have been successfully implemented to control mechanical vibrations \cite{jia2018}. Examples of ways of controlling vibrational modes interacting with emitters using phononic crystals have been reported for two-dimensional materials (operating in the MHz range) \cite{qian2023} and organic molecules (in the GHz) \cite{gurlek2021a}. Given that photosynthetic complexes of interest are often purified in solution, they can be deposited and integrated within phononic crystals without the requirement of a specific material system (unlike emitters that need to be grown in specific matrices or material stacks). This characteristic provides more flexibility in the design of phononic crystals in a material of choice, whose mechanical properties will allow to match the THz vibrational frequencies of interest for the specific emitter under study \cite{Sandeep2018}. 

Finally, in order to study the role of the vibrational resonance condition, an alternative approach is to instead alter site energies, which in turn shift the excitonic energies, and thus alter such a resonance. This can be achieved, for example, with either mutant protein complexes, or by oxidation of the bacteriochlorophyll \cite{higgins2021, higgins2021a}.

\section{Discussion}\label{sec:discussion}

Here we have studied how quantum coherent vibronic influence on exciton transport in single complex molecular emitters may be probed via the study by spectral filtering second-order photon correlations. We have studied a prototype photosynthetic heterodimer model under weak incoherent illumination, and have seen, indeed, that the behaviour of FFPCs is extremely sensitive to the presence of coherent coupling between excitons and vibrations that contribute to quantum transport under steady state conditions. We first observe that even broad-band, or unfiltered, correlations yield increased anti-bunching in the presence of coherent vibrational modes, and hence the appearance that the system acts more like a two-level emitter. At zero-delay, FFPC autocorrelations are observed to probe coherent transport via comparison of correlations that signal particular transfer pathways. This is possible as the frequency filter enables time information to be encoded in zero time values via uncertainty in time of detection. It is observed that coherent and incoherent environmental effects have drastically different signatures in the behaviour of zero-delay FFPC autocorrelations, making even zero-time FFPCs a sensitive witness of coherent environmental modes.

Turning to time dependence of FFPCs, we note that cross-correlations may show a time-asymmetry due to a swapping of the order of measured photons. We have performed a detailed numerical study of the signatures of coherent vibrational transport, which yields increased directional transport when the vibrational mode is near resonant with the exciton energy gap. We find such directional transport leads to increased FFPC time-asymmetry, and crucially, this is asymmetry is dominated by the coherent contribution to such transport. We show analytically that FFPC asymmetry is related to a violation of a QDB relation in a doubled Hilbert space, and study the violation of QDB in this model numerically to corroborate our results, finding that indeed the vibrational coupling induces a QDB violating channel.

We note that QDB violation in a system described by a Lindblad master equation  implies the presence of coherences in the steady state in the Hamiltonian eigenbasis \cite{Alicki1976, Kossakowski1977, Bratteli1978, roberts2021}. The violation of QDB offers an advantage for the function of light-harvesting complexes under continuous incoherent excitation, enabling a steady-state current of exciton transfer from high- to low-energy excitonic states. 
This leads us to hypothesise that a crucial role of NESS coherences, and indeed coherent contributions to exciton transfer processes, is in enabling QDB violation.

Our work therefore puts forward measurements of photon cross-correlation asymmetry  as a promising experimental tool to obtain insightful information on the quantum dynamical mechanisms and their unambiguous non-classical features underpinning quantum transport process in NESS of single complex emitters, from biomolecules, chemical sensing systems to solid-state systems such as quantum dots. Experimental measurements of the predicted optical correlation asymmetries are within reach with current detector technology.

\section{Methods}\label{sec:methods}

\subsection{Frequency-filtered photon correlations: Perturbative sensor method}\label{app:sensor}

The perturbative method allows an $n$-th order correlation function to be calculated exactly via $n$-th order perturbation theory. Evaluating Eq. \eqref{eq:FFPCdef} is then simplified to the following expression \cite{Holdaway2018}:
\begin{align}\label{eq:g2_components}
    g^{(2)}(\omega_1, \omega_2, \tau) = \frac{I_0(\tau) + I_1(\tau) + I_2(\tau)}{\langle n_1\rangle \langle n_2\rangle},
\end{align}
where,
\begin{align}
    &I_0(\tau) = e^{-\Gamma_2 \tau} \Tr[\rho_{1, 1}^{1, 1}] \\&
    I_1(\tau) = 2\epsilon \textrm{Im}\left[  \int_0^\tau dt e^{-\Gamma_2 (\tau - t/2) + i\omega_2 t} \Tr[a_2\rho_{1, 0}^{1, 1}(t) ] \right]\\&
    I_2(\tau) = -2 \epsilon^2 \textrm{Re} \Bigg[ \int_{0}^\tau dt_2 \int_0^{t_2} dt_1 e^{-\Gamma_2(\tau - \frac{t_1 + t_2}{2}) + i\omega_2 (t_2 - t_1)} \\& \qquad \qquad \times \Tr[a_2(t_2 - t_1)\rho_{1, 0}^{1, 0}(t_1) a_2^\dagger ] \Bigg]\nonumber
\end{align}
and for $\tau < 0$ indices 1 and 2 are swapped.
Here each term depends on a conditional density operator $\rho_{j_1, j_2}^{j_1^\prime, j_2^\prime}$, defined such that the steady state of the system and sensors, which are introduced as ancillary systems for calculation of FFPCs (see SM), is $\rho_{ss} = \sum_{j_1, j_2, j_1^\prime, j_2^\prime}\rho_{j_1, j_2}^{j_1^\prime, j_2^\prime} \otimes |j_1\rangle \langle j_1^\prime| \otimes |j_2\rangle \langle j_2^\prime$. $\rho_{j_1, j_2}^{j_1^\prime, j_2^\prime}$. $\rho_{j_1, j_2}^{j_1^\prime, j_2^\prime}$ is thus conditioned on the states $|j_1\rangle \langle j_i^\prime|$ of the sensor $i$.

$I_1$ and $I_2$ can be seen to contribute terms proportional to coherences in the conditional density operators, and thus we can expect these to have non-trivial contributions to the asymmetry. These are shown in Fig. \ref{fig:g2_components} of the SM, where we see that the dominant component, especially for long times, is $I_2$.
In Fig. \ref{fig:g2_components} of the SM we see that the asymmetry is dominated by $I_2$.
In the SM we further develop a numerically efficient approach via an algebraic solution to these integrals in the form of Sylvester equations \cite{Bhatia1997Jan}.

\subsection{Correlation asymmetry, microreversibility and quantum detailed balance}\label{sec:microrev_g2}

In \cite{carmichael1976} it was shown that the QDB as defined in Eq. \eqref{eq:qdb_def} is a consequence of microreversibility of dynamics under time reversal as defined above. Here we follow the same approach to show that symmetry of correlations similarly follow from a notion of time-reversal that includes swapping in the order of measurement events. The latter is precisely how we define negative time FFPCs in the above text. We here use the notation $\overleftarrow{a}$ to refer to this measurement order reversal combined with time reversal, such that $\overleftarrow{a}_1 = \tilde{a}_2$. 

We note that this can be written in terms of the (unitary) SWAP operator \cite{nielsen2010} which fulfills $S = S^{-1}$ and the complex conjugation operator as $T_S = SK$, and thus fulfills the general form for an anti-linear operator that must be fulfilled by a time-reversal symmetry as described above. We thus have that the  microreversibility condition under this time and measurement reversal implies $\tilde{a}_1 = a_1$, so $\overleftarrow{a}_1 = a_2$, and $\overleftarrow{U}(\tau) = U^\dagger(\tau)$. Notably, in the sensor method approach to FFPCs described below swapping the sensor operators swaps the central frequency of the sensor, and leaves the system evolution invariant. 

In Ref. \cite{denisov2002} it was shown for Bosonic emitters that the asymmetry of correlations can be understood in terms of a violation of detailed balance under the assumption $\tilde{a} = a$, where the detailed balance relation becomes 
\begin{align}\label{eq:qdb2}
    \langle A(t) B\rangle = \langle B^\dagger(t) A^\dagger\rangle.
\end{align}
It is this case of time-reversal invariant observables that we analyse here, and show the relation of correlation asymmetry to detailed balance follows for arbitrary emitters in a doubled Hilbert space.

We begin by defining the potentially asymmetric contribution to FFPCs, which are confined to the numerator, in terms of the probability of two successive measurements 
\begin{align}
    G_{12}^{(2)}(\tau) &= \langle a_1^\dagger a_2^\dagger(\tau) a_2(\tau) a_1 \rangle.
\end{align}
We note that FFPCs may be expressed in this form via sensor method approach \cite{DelValle2012, Holdaway2018} described below.
Now, assuming microreversible dynamics under a unitary operator $U(\tau) = e^{-iH\tau}$ we have
\begin{align}
    G_{12}^{(2)}(\tau) &= \Tr[a_1^\dagger U^\dagger(\tau) n_2(\tau) U(\tau) a_1 \rho] \nonumber \\&
    = \Tr[\overleftarrow{\rho} \overleftarrow{a_1}^\dagger (\overleftarrow{n_2(\tau)})^\dagger \overleftarrow{a_1}] \\&
    = \Tr[\overleftarrow{\rho} a_2^\dagger (\overleftarrow{U}^\dagger(\tau) n_1 \overleftarrow{U}(\tau))^\dagger a_2] \nonumber \\&
    = \Tr[\rho a_2^\dagger U^\dagger(\tau) n_1 U(\tau) a_2] \nonumber \\&
    = G^{(2)}_{21}(\tau) \nonumber
\end{align}
where in the second line we have exploited that $\Tr[\overleftarrow{O}^\dagger] = \Tr[O]$. We thus see that symmetry of $U$ under the SWAP operator leads to symmetric correlations. We can relate this symmetry to an expression of the form of Eq. \eqref{eq:qdb2} via rewriting $G_{12}(\tau)$ as 
\begin{align}\label{eq:begin}
    \Tr [ a_1^\dagger a_2^\dagger(\tau) a_2(\tau) a_1 \rho_{ss} ] & = \Tr [ a_2^\dagger(\tau) a_2(\tau) a_1 \rho_{ss}a_1^\dagger ] \nonumber \\& 
    = \Tr [ n_2(\tau) M_1[\rho] ],
\end{align}
where the map $M_1[\rho] : \mathcal{H}_a \to \mathcal{H}_a$ is defined by $M_1[\rho] = a_i \rho a_i^\dagger$ can be written in terms of the equivalent Choi state in a doubled Hilbert space $C^{(M_1)} \in \mathcal{H}_{s_1} \otimes \mathcal{H}_{s_2}$ (see SM \cite{SM} for a summary of the Choi-Jamio\l{}kowski isomorphism), where we now label the doubled system subspaces with indices 1 and 2 to avoid confusion, as
\begin{align}
     M_1[\rho] = \Tr_{s_1}[ (\rho_{s_1}^T \otimes \mathbb{1}_{s_2}) C^{(M_1)} ],
\end{align}
and thus
\begin{align}\label{eq:doubled_hilbert_g2}
    G^{(2)}_{12}(\tau) & =  \Tr_{s_2} [ n_2(\tau) \Tr_{s_1}[(\rho_s^T \otimes \mathbb{1}_s) C^{(M_1)} ]] \nonumber \\ & 
     =\Tr_{s_2} [ n_2(\tau) \Tr_{s_1}[\rho^\prime C^{(M_1)} ]] \\ & 
    =  \Tr [C^{(M_2)} n_1(\tau) \rho^\prime ]. \nonumber 
\end{align} negative time correlation is,
\begin{align}\label{eq:doubled_hilbert_g2_neg}
    G^{(2)}_{21}(\tau) & =  \Tr_{s_2} [ n_1(\tau) \Tr_{s_1}[\rho^\prime C^{(M_2)} ]] \\ & 
    =  \Tr [C^{(M_2)} n_1(\tau) \rho^\prime ]. \nonumber 
\end{align}

We finally use that $\Tr[\overleftarrow{O}^\dagger] = \Tr[O]$, and write, 
\begin{align}\label{eq:doubled_hilbert_g2_neg_2}
    G^{(2)}_{21}(\tau)  & =  \Tr [(\overleftarrow{C}^{(M_1)} \overleftarrow{n_2}(\tau) \overleftarrow{\rho}^\prime )^\dagger] \nonumber \\ &
    =\Tr[n_1(\tau)^\dagger C^{(M_2)\dagger} \rho^\prime ],
\end{align}
and thus if we enforce FFPC symmetry, $G^{(2)}_{12}(\tau) = G^{(2)}_{21}(\tau) $, we have
\begin{align}
     \Tr [ n_2(\tau) C^{(M_1)} \rho^\prime ]  =  \Tr [C^{(M_1)\dagger} n_2(\tau)^\dagger \rho^\prime ],
\end{align}
which is of the same form as the statement of QDB in \eqref{eq:qdb2}.

Finally, as the denominators of $g^{(2)}(\omega_1, \omega_2, \tau)$ are trivially time symmetric, we have that the same property is applicable to the normalised form of the FFPC used in the main text.

\section{Data Availability}

Python code for sensor method calculations of FFPCs via QuTip \cite{johansson2011, johansson2013} is available at \cite{github}. Data available from the corresponding authors upon reasonable request.

\section{Acknowledgements} 
We thank the Engineering and Physical Sciences Research Council (EPSRC UK) and the Gordon and Betty Moore Foundation grant GBMF8820 for financial support. We thank David I. H. Holdaway for useful discussions in the early stages of this work. C. N thanks S. Vezzoli for useful discussions. A.O-C thanks Graham Fleming , Quanwei Li,  K. Birgitta Whaley and Robert L Cook for discussions. The authors acknowledge use of the UCL Myriad High Performance Computing Facility.

\section{References}
\bibliographystyle{apsrev4-1}
\bibliography{bibliog}

\begin{widetext}
\begin{center}
\begin{titlepage}
  \centering
  \vskip 60pt
  \LARGE  Supplemental Material: \\ Witnessing coherent environmental enhancement of quantum transport via time-asymmetry of two-colour photon correlations\par
  \vskip 1.5em
  \large Charlie Nation$^{1}$, Valentina Notararigo$^{1}$, Hallmann Gestsson$^{1}$, Luca Sapienza$^{2}$, and Alexandra Olaya-Castro$^{1}$ \par
  \vskip 1.0em
\small \textit{$^{1}$Department of Physics and Astronomy, University College London, Gower Street, WC1E 6BT, London, United Kingdom} \\
\small \textit{$^{2}$Department of Engineering, University of Cambridge, Cambridge, CB3 0FA, United Kingdom} \\
  \vskip 0.5em

  \today
    \vskip 2.5em

\end{titlepage}
\end{center}
\setcounter{equation}{0}
\setcounter{figure}{0}
\setcounter{table}{0}
\setcounter{page}{1}
\setcounter{section}{0}

\makeatletter
\renewcommand{\theequation}{S\arabic{equation}}
\renewcommand{\thefigure}{S\arabic{figure}}
\renewcommand{\bibnumfmt}[1]{[S#1]}
\renewcommand{\citenumfont}[1]{S#1}

\section{Model details}\label{app:model}

In this section we provide additional details and numerical results focusing on properties of the studied dimer model.

\subsection{Steady state coherences}

In this section we show directly how the coherences in the steady state change with pump power $P_{X_1}$, vibrational coupling $g_v$, and vibrational frequency $\omega_v$ for each dimer model. This is shown in Fig. \ref{fig:coherences}, where we show in a), c) the excitonic coherence, and in b) d) the Klyshco parameter \cite{klyshko1996} defined by $B_1=P_0P_1 - P_0^2$ for the modulation of populations of nearest vibrational states, where $P_n=\textrm{Tr}[|n\rangle\langle n| {\rho}_{ss} ]$ and c) exciton-vibration coherence. We observe, as expected, that increasing $V$ and $g_v$ leads to an increase in excitonic and vibrational coherences, respectively. The increase of electronic coherence for increasing $g_v$, however, is non-monotonic, with a minima for intermediate values. Similarly, in Fig. \ref{fig:coherences_omegas} we show the dependence of coherence measures on the vibrational mode frequency $\omega_v$, observing that the Klyshco parameter indicates larger quantum behaviour for lower vibrational energies. Notably, we observe that the excitonic coherence in Fig. \ref{fig:coherences_omegas}c), that closest to biological conditions, show minima around the vibrational resonance conditions with the excitonic energy gap. This corresponds to regions where QDB is minimally violated at long times, and maximal regions of QDB violation, such as $\omega_v \sim 800$ cm$^{-1}$ show maximal excitonic coherence in the steady state. 
\begin{figure}
    \includegraphics[width=0.48\textwidth]{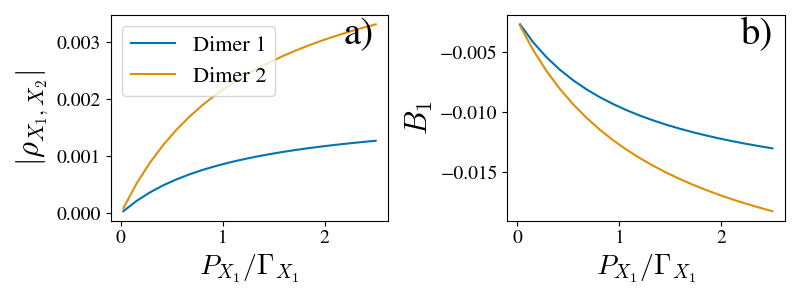} \\    \includegraphics[width=0.48\textwidth]{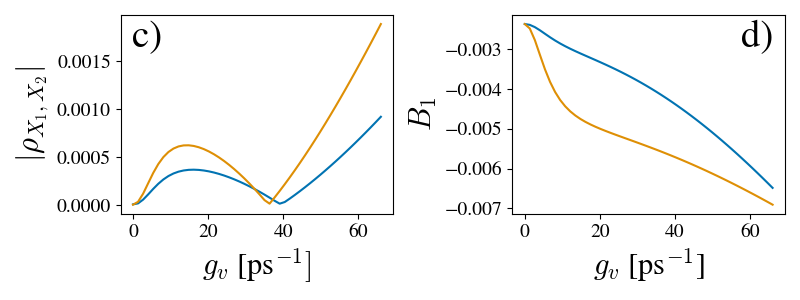}
    \caption{System coherence measures. a), c) show excitonic coherence $\rho_{|X_1\rangle \langle X_2|} = \Tr[|X_1\rangle \langle X_2| \rho_{ss} ]$ and their dependence on pump power $P_{X_1}$ and vibrational coupling $g_v$ respectively. b), d) show the respective dependence's of the Klyshco parameter (see text), the negativity of which is a measure of vibrational coherence.}
    \label{fig:coherences}
\end{figure}
We thus see, as noted in the main text, that parameter regimes more dominated by unitary dynamics do not necessarily correspond to regions of large steady-state coherence. We further note that QDB violation implies the existence of such coherences, thus we may expect them to be larger when QDB is maximally violated (though this relation is not necessarily monotonic). 

\begin{figure}
    \includegraphics[width=0.48\textwidth]{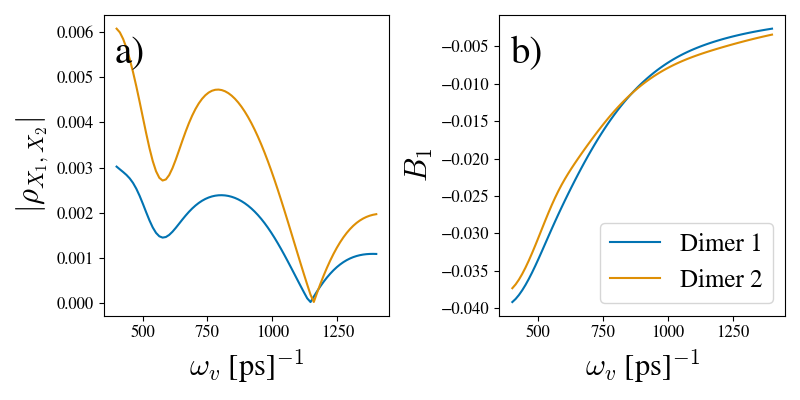}
    \caption{System coherence measures and their dependence on vibrational mode frequency $\omega_v$. a)shows excitonic coherence $\rho_{|X_1\rangle \langle X_2|} = \Tr[|X_1\rangle \langle X_2| \rho_{ss} ]$ for $g_v = 267.1$ cm$^{-1}$. b) shows the respective dependence of the Klyshco parameter (see text), the negativity of which is a measure of vibrational coherence.}
    \label{fig:coherences_omegas}
\end{figure}

In Fig. \ref{fig:dynamics} we show the dynamics exciton populations and coherences, starting from the initial state $\rho(0) = |X_1\rangle \langle X_1| \otimes \rho_\beta$, where $\rho_\beta = \frac{1}{Z}e^{-\beta H_{vib}}$ is the thermal state of the vibrational modes. Here we see that, as expected, dimer 2 has increased coherences and faster transport due to increased excitonic delocalisation. We also see clearly that the presence of the vibrational mode acts to speed up transport, and increase excitonic coherence.

\begin{figure}
    \includegraphics[width=0.48\textwidth]{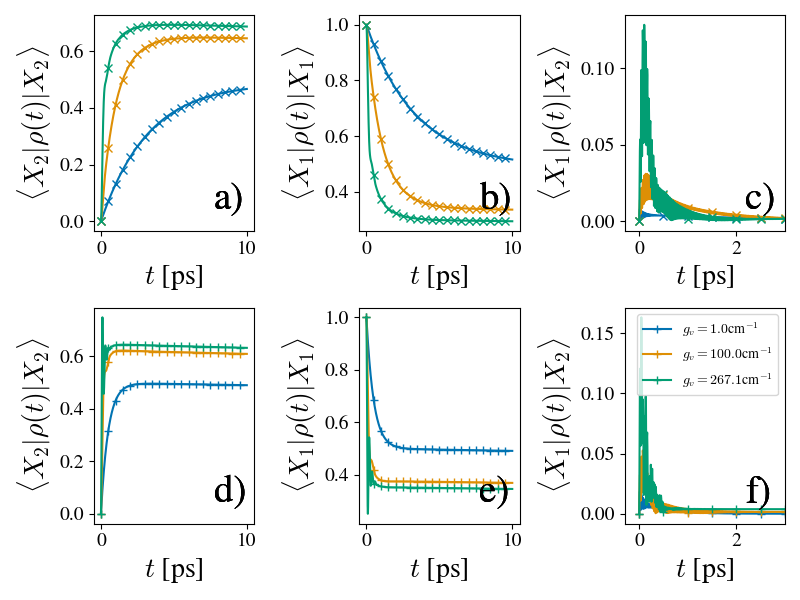}
    \caption{Exciton population and coherence dynamics for a-c) dimer 1 and d-f) dimer 2.}
    \label{fig:dynamics}
\end{figure}
\subsection{Role of the center of mass mode}\label{app:com}

\begin{figure}
    \includegraphics[width=0.98\linewidth]{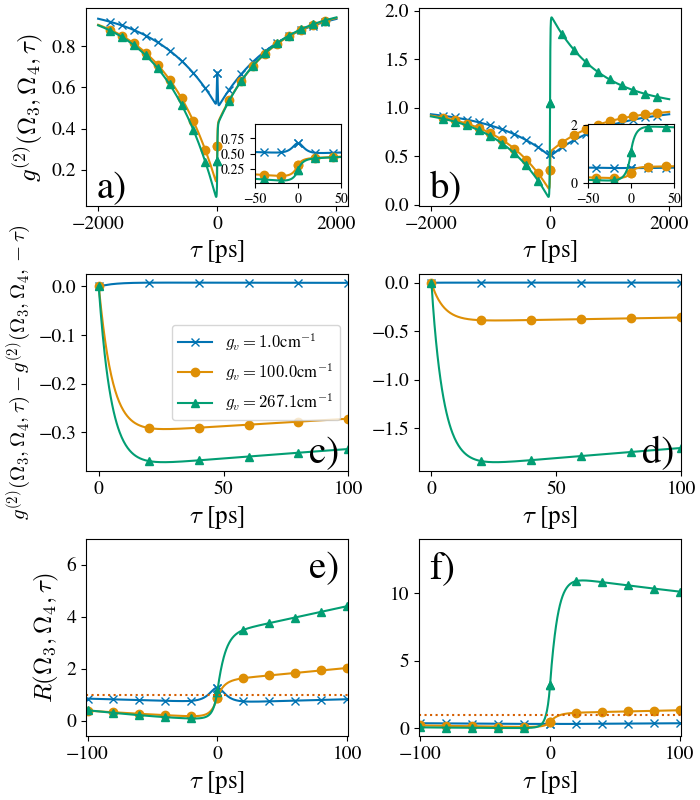}
    \caption{As in Fig \ref{fig:g2t}, however with no vibrational COM mode. We see here a monotonic increase in asymmetry with $g_v$, as expected in this case as the only effect of $g_v$ is via the coherent transfer mechanism.}
    \label{fig:no_com}
\end{figure}

As can be seen from Eq. \eqref{H_diag}, the center of mass (COM) mode only couples to the electronic degrees of freedom via the operator $M$, which does not couple excitonic states, and, further, in the single excitation subspace is simply the identity. Thus, in cases where only dynamics in this subspace is considered, the COM mode may be ignored altogether \cite{Oreilly2014}. However, in the presence of emission and pump processes incoherently coupling the electronic ground and excited states, the COM mode coupling is no-longer proportional to the identity, and population of this mode can be seen to cause an effective renormalisation of the excited state energies proportional to the COM mode displacement.

In Fig. \ref{fig:no_com} we show an analogous calculation to that in Fig. \ref{fig:g2t}, however with no COM mode. We observe that there is a substantial quantitative change in the results, however they follow the same qualitative behaviour. This is because whilst the COM mode indeed alters the dynamics of the system, it's role is essentially in causing dynamical disorder in the energy gap between ground and excited states, and does not alter the mechanism of QDB violation.

\subsection{Dependence on pure dephasing}

\begin{figure}
    \includegraphics[width=0.98\linewidth]{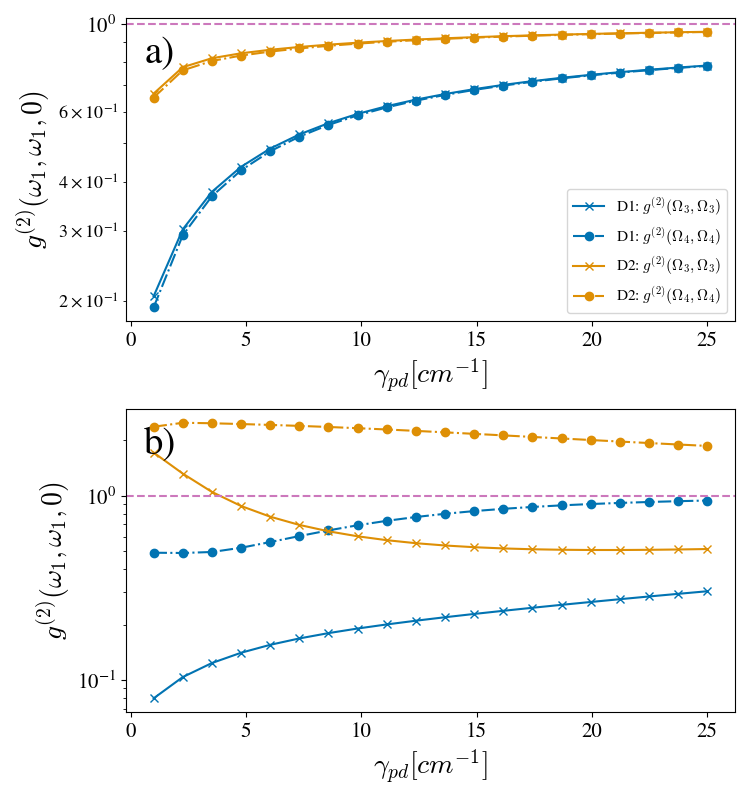}
    \caption{Photon autocorrelations as in Fig \ref{fig:g20}c,  however for varying $\gamma_{pd}$ rather than varying $g_v$. a) Shows dependence when no coherent mode is present ($g_v = 0$), and b) shows $g_v = 267.1$ cm$^{-1}$.}
    \label{fig:g20_gpd}
\end{figure}

In this section we provide additional numerics which demonstrates the role of pure dephasing in the dimer model. Pure dephasing, as discussed in the main text, enables an incoherent exciton transfer mechanism that relies on the presence of delocalised excitons, and thus may have some overlapping characteristics with $g_v$, which enables coherent transfer.

In Fig. \ref{fig:g20_gpd} we show the dependence of auto-correlations taken around the exciton energies $\Omega_3$, $\Omega_4$ on $\gamma_{pd}$. In the main text we say that these autocorrelations drastically changed with $g_v$, and, being identical for $g_v = 0$, as the coupling was increased we saw that due to differing dependencies on excitonic transfer pathways the auto-correlations differed for large $g_v$, with the preferred pathway becoming more bunched, and the other becoming more antibunched. Here we see that similarly observing the auto-correlations and changing $\gamma_{pd}$, and hence the incoherent transfer rate, we observe a change in each auto-correlation. However, for $g_v = 0$, Fig. \ref{fig:g20_gpd}a), we see that the autocorrelations are almost identical for all values of $\gamma_{pd}$. For finite $g_v$, Fig. \ref{fig:g20_gpd}b), we see that the autocorrelations indeed differ in the manner described, however as $\gamma_{pd}$ is increased the autocorrelations move together. We thus confirm that, indeed, the core mechanism leading to deviations in autocorrelations is the coherent transfer pathway indicated in Fig. \ref{fig:g20_pathways}b) of the main text.

\begin{figure}
    \includegraphics[width=0.98\linewidth]{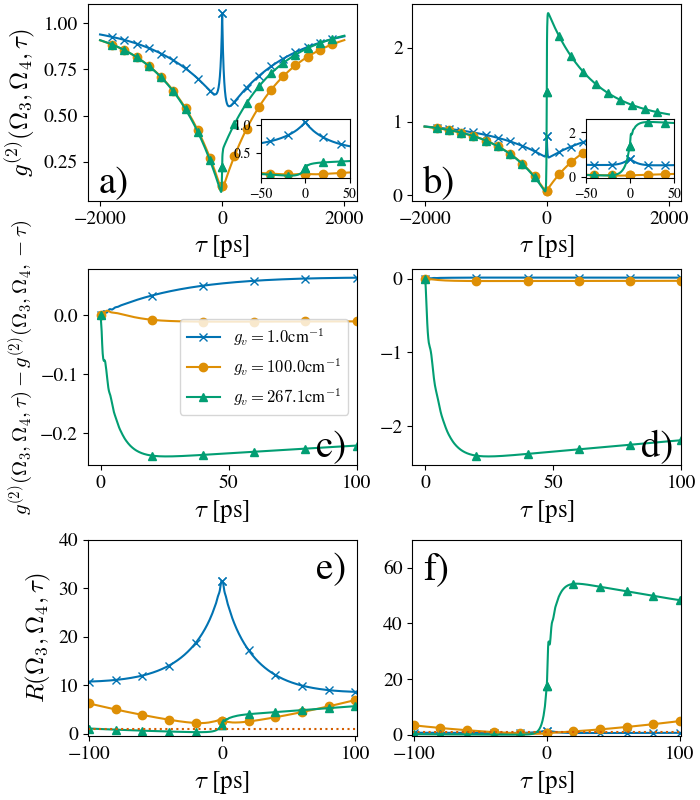}
    \caption{As in Fig \ref{fig:g2t}, with $\gamma_{pd} = 1$ ps$^{-1}$. Here we see that the reduced dephasing leads to an increased sensitivity to $g_v$. The COM mode here plays similarly important role to Fig \ref{fig:g2t}, leading to a non-monotonic increase in asymmetry with $g_v$.}
    \label{fig:g2t_gpd1}
\end{figure}

In Fig. \ref{fig:g2t} we show the time dependence of FFPC cross-correlations for $\gamma_{pd} = 1$ ps$^{-1}$. We observe for weak $g_v$ a very similar behavior as with $\gamma_{pd} = 10$ ps$^{-1}$ shown in Fig. \ref{fig:g2t} of the main text, that is, we have little asymmetry, and the CSI is violated in a near symmetric manner. Increasing $g_v$ we observe again the clear directional quantum behaviour, with the CSI violation in the positive time direction being extended to very long times. We note that for $\gamma_{pd} = 1$ cm$^{-1}$ we observe a non-monotonic increase in asymmetry with $g_v$, which we associate to the effective noise of the COM mode which also increases with $g_v$. Indeed, if we only include the RD mode (not shown) this dependence is once more monotonic.

Our results here thus corroborate the results of the main text: that photon correlation asymmetry is sensitive to coherent exciton transfer.

\subsection{Derivation of emission jump operators}\label{app:jump_ops}

In this section we derive the form of the jump operators contributing to the emission processes in the Gorini, Kossakowski, Lindblad and Sudarshan (GKSL) master equation used in the main text. We use a standard quantum optical master equation \cite{BP_c3}, starting from the light matter interaction Hamiltonian for an $N$-site Frenkel exciton model,
\begin{align}
    H_I = \sum_m \hat{\mu}_m \otimes \hat{B}_m,
\end{align}
with $\hat{B}_m = g_m( \hat{b}^\dagger_m + \hat{b}_m)$ proportional to displacement operator of the electromagnetic field modes coupled to site $m$ through the dipole operator, which may be expressed in the Hamiltonian eigenbasis as follows;
\begin{align}\label{eq:A_omega_def}
    \hat{\mu}_m &= \mu_m (|m\rangle \langle g| + |g\rangle \langle m|) \nonumber \\&
    = \mu_m \sum_{\nu, \nu^\prime} |F_\nu\rangle \langle F_\nu | (|m\rangle \langle g| + |g\rangle \langle m|)|F_{\nu^\prime}\rangle \langle F_{\nu^\prime} | \\&
    = \sum_{\omega} \mu_m(\omega) \hat{\Pi}(\omega)  \nonumber  \\&
    = \sum_{\omega} \hat{A}_m(\omega) \nonumber
\end{align}
where we have defined $\mu_m(\omega) := \mu_m \langle F_\nu | (|m\rangle \langle g| + |g\rangle \langle m|)|F_{\nu^\prime}\rangle$ and $\hat{\Pi}(\omega) =|F_\nu\rangle \langle F_{\nu^\prime} |$, with $\omega = E_\nu - E_{\nu^\prime}$ as the energy difference of the transition $|F_\nu \rangle \to |F_{\nu^\prime}\rangle$. We note that whilst $\hat{\mu}_m$ is Hermitian, the components $\hat{A}_m(\omega)$ are not in general, and that as $|F_\nu\rangle$ are eigenstates of $H$ with energy $E_\nu$, we have
\begin{align}
    [H, \hat{A}(\omega)] = - \omega \hat{A}(\omega), \,  [H, \hat{A}(-\omega)] = \omega \hat{A}(\omega).
\end{align}

Then, after the Born-Markov and rotating wave approximations \cite{BP_c3}, we can write the GKSL form dissipator due to this interaction in the form,
\begin{align}\label{eq:qo_me}
    \mathcal{D}[\rho] = \sum_{\omega}\sum_{m, m^\prime} \gamma_{m, m^\prime}(\omega) (\hat{A}_m(\omega) &\rho \hat{A}^\dagger_{m^\prime}(\omega) \\& \nonumber + \{\hat{A}^\dagger_{m^\prime}(\omega) \hat{A}_m(\omega) , \rho\}),
\end{align}
with 
\begin{align}\label{eq:gamma_def}
    \gamma_{m, m^\prime}(\omega) = \int_{-\infty}^\infty ds e^{i\omega s} \langle \hat{B}^\dagger_m(s) \hat{B}_m(0) \rangle.
\end{align}
Following Ref. \cite{BP_c3} (see sec 3.4), and assuming the electromagnetic environment to be in the limit of small photon number $N(\omega) \ll 1$, we have
\begin{align}\label{eq:qo_me2}
    \mathcal{D}_m[\rho] = \sum_{\omega>0} \gamma_{m} (\hat{A}_m(\omega) \rho \hat{A}^\dagger_{m}(\omega) + \{\hat{A}^\dagger_{m}(\omega) \hat{A}_m(\omega) , \rho\}),
\end{align}

As we will see, the nature of the dipole operator, enabling transitions from the ground to excited states in the site basis, further restricts the contributing transitions $\Pi(\omega)$.

Before continuing, we note some important features of the relations between the three relevant bases in which processes are here described. First, the site basis $(\{ m \} \, | \, m\in [0, N]) \in \mathcal{B}(\mathcal{H}_{el})$, with $|m=0\rangle = |g\rangle$ being the ground state. The excitonic basis diagonalises the electronic Hamiltonian, which does not couple ground ($m=0$) and excited ($m>0$) states, thus we have similar form of the excitonic basis $(\{ |X_i\rangle \} \, | \, i \in [0, N]) \in \mathcal{B}(\mathcal{H}_{el})$, with $|X_0\rangle = |m=0\rangle = |g\rangle$ labelling the ground state. Finally, we have the total Hamiltonian (vibronic) eigenstates $|F_\nu\rangle$ which diagonalise the system Hamiltonian including the coherent vibrational modes. We truncate each local vibrational mode at an occupation number $L$, and thus the total number of vibrational energy levels is $L^2$, and $\nu \in [0, (N + 1) L^2]$ for a Frenkel exciton model with only a single excitation (we will discuss the second excited state contribution below). We further note, as the vibrational modes do no act to couple the electronic excited states to the ground state, the first $L^2$ vibronic eigenstates have the simple form $|g, l\rangle$, where $l$ is a multi-index determining the vibrational levels of both modes.

Thus, we have
\begin{subequations}
\begin{align}
    &|F_\nu\rangle = \sum_{i, l} c_{i, l}(\nu) |X_i, l\rangle, \\&
    |X_i \rangle = \sum_m a_m(i) |m \rangle,
\end{align}
\end{subequations}

with $c_{il}(\nu) = \langle X_i, l| F_\nu \rangle$ and $a_m(i) = \langle m | X_i \rangle$, with ground state components $c_{0l}(\nu) = \langle g, l| F_\nu \rangle = \delta_{\nu, (0, l)}$ and $a_0(i) = \langle g | X_i \rangle = \delta_{i0}$ respectively. 
Thus, to evaluate the components in Eq. \eqref{eq:A_omega_def}, we can write
\begin{align}
    \langle F_\nu | m \rangle & = \sum_{il} c_{il}(\nu) \langle X_i, l | m\rangle \nonumber \\& 
    =  \sum_{il} c^*_{il}(\nu) a^*_{m}(i) \langle l| 
\end{align}
and
\begin{align}
    \langle F_\nu | g \rangle &= \sum_{il} c_{il}^*(\nu) \langle X_i, l | g\rangle \nonumber \\& 
    = \sum_l  \delta_{\nu, (0, l)} \langle l |
\end{align}
such that
\begin{align}
    \hat{A}_m(\omega) &= \mu_m \langle F_\nu | \left(|m\rangle \langle g| + |g\rangle \langle m|\right)|F_{\nu^\prime}\rangle \hat{\Pi}(\omega)  \nonumber \\&
    =  \mu_m (\langle F_\nu |m\rangle \langle g |F_{\nu^\prime} \rangle + \langle F_\nu|g\rangle \langle m|F_{\nu^\prime} \rangle )\hat{\Pi}(\omega)  \\&
        =  \mu_m (  \sum_{il} c^*_{il}(\nu) a^*_{m}(i) \langle l|  \sum_{l^\prime}  \delta_{\nu^\prime, (0, l^\prime)} |l^\prime \rangle \nonumber \\&
        \qquad \qquad +  \sum_{l}  \delta_{\nu, (0, l)} \langle l | \sum_{il^\prime} c_{il^\prime}(\nu^\prime) a_{m}(i) | l^\prime \rangle  ) \hat{\Pi}(\omega) \nonumber \\&
     =  \mu_m \sum_{il} (  c^*_{il}(\nu) a^*_{m}(i) \delta_{\nu^\prime, (0, l)} + \delta_{\nu, (0, l)}  c_{il}(\nu^\prime) a_{m}(i)  )\hat{\Pi}(\omega) \nonumber 
\end{align}
We thus immediately see that the possible $\omega$ values are those that are close to the ground-excited state transition energy. Concretely, as $\hat{\Pi}(\omega) = (|F_{\nu}\rangle \langle F_{\nu^\prime}|)_{E_\nu - E_\nu^\prime = \omega}$ we have, defining $\hat{\sigma}_{\nu l} := |g, l\rangle \langle F_\nu|$ and  $F_{m l \nu} := \mu_m \sum_i c_{il}(\nu) a_m(i)$,
\begin{align}
    \hat{A}_m(\omega) = \sum_{l} F_{m l \nu} \hat{\sigma}_{\nu l} 
\end{align}
and
\begin{align}
    \hat{A}_m(-\omega) = \hat{A}^\dagger_m(\omega) = \sum_{l} F^*_{m l \nu} \hat{\sigma}_{\nu l}^\dagger 
\end{align}
Thus, we obtain an expression in terms of weighted transitions from Hamiltonian eigenstates to electronic ground states with local vibrational excitations. We assume constant dipole moments $\mu_m = \mu$, and thus write a single rate $\gamma$ for each transition that absorbs the dipole moment contribution to the weights, and define
\begin{align}\label{eq:Aomega}
    A(\omega) &= \sum_m A_m(\omega)  \nonumber \\ &
    = \mu \sum_{m} \sum_i \sum_l c_{il}(\nu) a_m(i) \hat{\sigma}_{vl}.
\end{align}
The operators are then weighted by the eigenstate components mixing site, excitonic, and vibronic bases. There are thus $2L + L$ jump operators for $L$ vibrational levels ($2L$ values of $\nu$, $L$ values of $l$).

In practise it is not convenient to label our set of jump operators by a frequency $\omega$, but rather in terms of quantum numbers that label the unique jump operators. We see in Eq. \eqref{eq:Aomega} that the operators are defined by the indices $\nu, l$, and thus we relabel
\begin{align}
    \sum_\omega A(\omega) = \sum_{\nu} \sum_l F_{\nu, l} \sigma_{\nu, l},
\end{align}
where 
\begin{align}
    F_{\nu, l} &= \sum_m \sum_i c_{il}(\nu) a_m(i) \nonumber \\&
    = \sum_m \sum_i \langle X_i , l | F_\nu\rangle \langle m| X_i\rangle \\&
    = \sum_m \langle m, l | F_\nu\rangle.\nonumber
\end{align}

\begin{figure}
    \includegraphics[width=0.95\linewidth]{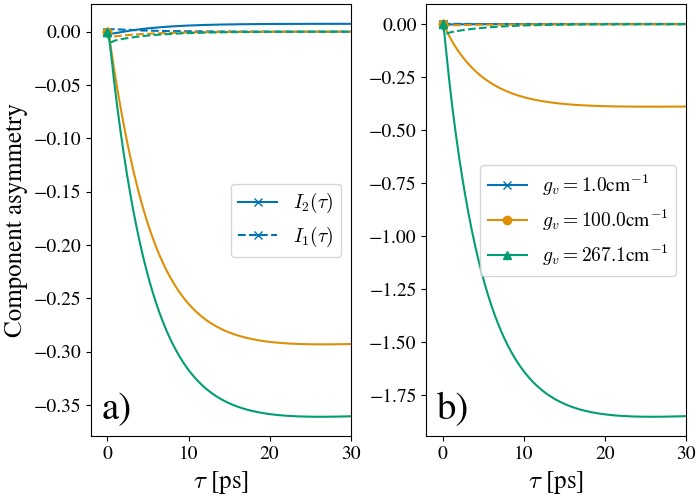} 
    \caption{Time asymmetry of components of $g^{(2)}(\omega_1, \omega_2, \tau)$. For dimer 1 (a)) and dimer 2 (b)) corresponding to first 100 ps of Fig. \ref{fig:g2t} a), b), respectively. $I_0$ is trivially symmetric so not shown. $I_1$ asymmetry is shown in the dashed lines, and is significantly smaller than the dominant contribution to asymmetrical photon correlations, which is from $I_2$, shown in solid lines.}
    \label{fig:g2_components}
\end{figure}

We can extend the above discussion to include the second excited state noting the additional contribution from $\hat{\mu}_{m, m^\prime} = \mu_{m^\prime} |m, 0\rangle \langle m, m^\prime| + \mu_{m} |m^\prime, m \rangle \langle m, 0| $ for $m \neq m^\prime$. For the dimer model discussed, then, we have two additional processes due to the transitions $|1, 2\rangle \to |m\rangle $ for $m \in [1, 2]$, which have the associated jump operators 
\begin{align}
    \hat{A}_{12 \to m}(\omega) &= \langle F_\nu |m\rangle \langle 1, 2 |F_{\nu^\prime}\rangle \hat{\Pi}(\omega) \nonumber \\&
    = \sum_{i, l} c^*_{i, l}(\nu) a^*_m(i) \delta_{\nu, (1, 2, l)} \hat{\Pi}(\omega) \\&
    = \sum_{i, l} c^*_{i, l}(\nu) a^*_m(i) |X_i\rangle \langle 1, 2|,\nonumber
\end{align}
where we have used that similarly to the ground state contribution the vibrational modes do not mix the single and doubly excited manifolds, and thus $\langle F_\nu| 1, 2\rangle = \delta_{\nu, (1, 2, l)}\langle l|$.

\section{Frequency-filtered photon correlations}\label{SM:sensor_method}

In the above we exploit a perturbative approach to the calculation of FFPCs \cite{Holdaway2018}, based on the sensor formalism of Ref. \cite{DelValle2012}. This approach relates the filters around frequency $\omega_m$ of linewidth $\Gamma_m$ to a hypothetical sensor, made up of a two-level system of the same frequency, with annihilation (creation) operator $\zeta_m$ ($\zeta^\dagger_m$), with a Markovian decay channel at a rate given by the frequency linewidth $\Gamma_m$. The sensors are coupled to the operator mode $a_m$ for which the photon correlations are to be measured via
\begin{align}
H_{I,m} = \epsilon (a_m \zeta^\dagger_m + a_m^\dagger \zeta_m).
\end{align}
We take $a_m = a = \sigma_{X_1} + \sigma_{X_2}$ in the above model.

The power spectrum of Fig. \ref{fig:diagram}b), c) is obtained via
\begin{align}\label{eq:power_spec}
S^{(1)}_{\Gamma_1}(\omega_1) = \frac{\Gamma}{2\pi\epsilon^2}\Tr [n_1 \rho_{ss}],
\end{align}
where $\rho_{ss}$ is the steady state of the total molecule-cavity-sensor system, and $n_m = \zeta_m^\dagger \zeta_m$. Similarly, for the filtered two-photon correlations, we have \cite{DelValle2012, Gonzalez-Tudela2013a, Holdaway2018}
\begin{align}\label{eq:g2tau_filtered}
g^{(2)}(\omega_1, \omega_2, \tau) = \frac{S^{(2)}_{\Gamma_1, \Gamma_2}(\omega_1, \omega_2, \tau)}{S^{(1)}_{\Gamma_1}(\omega_1) S^{(1)}_{\Gamma_2}(\omega_2)},
\end{align}
where,
\begin{align}
S^{(2)}_{\Gamma_1, \Gamma_2}(\omega_1, \omega_2, \tau) = \frac{\Gamma_1 \Gamma_2}{(2\pi \epsilon^2)^2}\Tr [\rho_{ss} \zeta_1^\dagger \zeta_2^\dagger(\tau)\zeta_2(\tau) \zeta_1].
\end{align}

The perturbative method allows an $n$-th order correlation function to be calculated exactly via $n$-th order perturbation theory, Eq. \eqref{eq:FFPCdef} is then \cite{Holdaway2018}:
\begin{align}\label{eq:g2_components_SM}
    g^{(2)}(\omega_1, \omega_2, \tau) = \frac{I_0(\tau) + I_1(\tau) + I_2(\tau)}{\langle n_1\rangle \langle n_2\rangle}.
\end{align}
Here each term depends on a conditional density operator $\rho_{j_1, j_2}^{j_1^\prime, j_2^\prime}$, defined such that the total steady state of the system and sensors is $\rho_{ss} = \sum_{j_1, j_2, j_1^\prime, j_2^\prime}\rho_{j_1, j_2}^{j_1^\prime, j_2^\prime} \otimes |j_1\rangle \langle j_1^\prime| \otimes |j_2\rangle \langle j_2^\prime$. $\rho_{j_1, j_2}^{j_1^\prime, j_2^\prime}$ is thus conditioned on the states $|j_1\rangle \langle j_i^\prime|$ of the sensor $i$. We then have for $\tau > 0$ \cite{Holdaway2018}
\begin{align}
    &I_0(\tau) = e^{-\Gamma_2 \tau} \Tr[\rho_{1, 1}^{1, 1}] \\&
    I_1(\tau) = 2\epsilon \textrm{Im}\left[  \int_0^\tau dt e^{-\Gamma_2 (\tau - t/2) + i\omega_2 t} \Tr[a_2\rho_{1, 0}^{1, 1}(t) ] \right]\\&
    I_2(\tau) = -2 \epsilon^2 \textrm{Re} \Bigg[ \int_{0}^\tau dt_2 \int_0^{t_2} dt_1 e^{-\Gamma_2(\tau - \frac{t_1 + t_2}{2}) + i\omega_2 (t_2 - t_1)} \\& \qquad \qquad \times \Tr[a_2(t_2 - t_1)\rho_{1, 0}^{1, 0}(t_1) a_2^\dagger ] \Bigg]\nonumber
\end{align}

The above perturbative sensor method has two numerical advantages over direct application of the sensor approach in Ref. \cite{DelValle2012}: first, the sensor coupling $\epsilon$ does not need fine tuning, as it vanished algebraically in determination of correlation functions, second the sensor Hilbert space does not need to be simulated, reducing the numerical complexity significantly for large systems such as the vibronic dimer studied above. The core drawback, however, is determination of the integral $I_2(\tau)$. Below we describe a more numerically efficient approach via an algebraic solution to these integrals.

\subsection{Components of FFPCs}

The three terms in Eq. \eqref{eq:g2_components} of the main text contribute different physical aspects of the photon correlations. For example, $I_1(\tau)$ simply contributes the zero-delay value, and decays at a rate set by the filter linewidth. We note that the  time-uncertainty of detection implied by the frequency resolution also contributes to the zero-delay value, and thus $I_1$ is very sensitive to experimental measurement parameters in both its zero-delay value and time dependence. As we have chosen $\Gamma_1 = \Gamma_2$, such that the frequency-filter linewidth is the same for both filters, the time-asymmetry of $I_0$ vanishes.

In Fig. \ref{fig:g2_components} we show the time dependence of the individual components of the FFPCs for each dimer model. We see that, as discussed in the main text, $I_2$ is the dominant contribution to long-time dynamics and asymmetries in FFPCs. 

\subsection{Efficient approach to sensor method calculations}\label{sec:improved_sensor}

Notice that whilst the sensor method is significantly more numerically tractable than direct integration of the four time- and normal-ordered integrals of Eq. \eqref{eq:g2tau_filtered}, $I_2(\tau)$ is nonetheless a two dimensional integral that we find is highly oscillatory (see Fig. \ref{fig:g2_components} of the SM), and thus remains numerically challenging for long timescales. We thus develop below an algebraic approach to the integrals in both $I_1(\tau)$ and $I_2(\tau)$.
First we write $I_1(\tau)$ in a Liouville space notation in which we may re-write the Hilbert-Schmidt norm as $\Tr[AB] = \llangle A^\dagger | B \rrangle$, such that
\begin{align}
    I_1(\tau) = 2\epsilon\textrm{Im}[\bbraket{a_2^\dagger |G_1(\tau)|\,\rho^{1, 0}_{1, 1}(0)}],
\end{align}
where 
\begin{align}
    G_1(\tau) = e^{-\Gamma_2 \tau} \int_0^\tau dt e^{(\frac{\Gamma_2}{2} + i\omega_2 + L)t}.
\end{align}
Here we used $\Tr[a_2 e^{\mathcal{L}t}[\rho^{1, 0}_{1, 1}(0)]] = \llangle a_1^\dagger| e^{Lt}|\,\rho^{1, 0}_{1, 1}(0)\rrangle$ where $L$ is the Liouville space representation of $\mathcal{L}$ such that we have $\kket{\mathcal{L}[\rho]} = L\kket{\rho}$. We may then directly integrate to obtain
\begin{align}
  G_1(\tau) = \frac{1}{\frac{\Gamma_2}{2} + i\omega_2 + L}(e^{(\frac{\Gamma_2}{2} + i\omega_2 + L)\tau} - e^{-\Gamma_2 \tau} \mathbb{1} ).
\end{align}
Thus, $I_1(\tau)$ may be obtained by calculating simply the reciprocal of a superoperator $\frac{\Gamma_2}{2} + i\omega_2 + L$, and evolving the state $|\rho^{1, 0}_{1, 1}(\tau)\rrangle$ in time.

A similar approach is obtained for $I_2$, where we first re-write the trace in the Liouville notation as 
\begin{align}
    \Tr[a_2(t_2 - t_1)\rho^{1, 0}_{1, 0}(t_1) a_2^\dagger ] & = \Tr[e^{\mathcal{L}^\dagger (t_2 - t_1)}[a_2] e^{\mathcal{L}t_1} [\rho^{1, 0}_{1, 0}] a_2^\dagger ] \nonumber \\  &
    = \Tr[a_2 e^{\mathcal{L}(t_2 - t_1)} [e^{\mathcal{L}t_1}[\rho^{1, 0}_{1, 0}]a_2^\dagger ]] \\&
    = \llangle a_2^\dagger |e^{L(t_2 - t_1)} (a_2^\dagger)^T_L e^{Lt_1} |\,\rho^{1, 0}_{1, 0}\rrangle, \nonumber
\end{align}
where in the second line we have exploited the quantum regression theorem, and in the last line we have used that $|ABC\rrangle = (C^T \otimes A)|B\rrangle$ for operators $A,\, B\, C$, and we define $(A)_L = A \otimes \mathbb{1}$,  $(A)_R = \mathbb{1} \otimes A$ as the left (right) acting superoperators. We can now re-write the integral for $I_2(\tau)$ as 
\begin{align}
    e^{-\Gamma_2 \tau} \int_0^\tau dt_2 \int_0^{t_2} dt_1  & e^{(\frac{\Gamma_2}{2} + i\omega_2)t_2 + (\frac{\Gamma_2}{2} - i\omega_2)t_1 }  \llangle a_2^\dagger |e^{L(t_2 - t_1)} (a_2^\dagger)^T_L e^{Lt_1} |\,\rho^{1, 0}_{1, 0}\rrangle\nonumber \\ & =  \llangle a_2^\dagger |G_2(\tau) |\,\rho^{1, 0}_{1, 0}\rrangle,
\end{align}
where
\begin{align}
    G_2(\tau) & = e^{-\Gamma_2 \tau} \int_0^\tau dt_2 \int_0^{t_2} dt_1 e^{(\frac{\Gamma_2}{2} + i\omega_2 + L)t_2 + (\frac{\Gamma_2}{2} - i\omega_2 + L)t_1 }  (a_2^\dagger)^T_L e^{Lt_1}
\end{align}
It is possible to perform the integral with respect to $t_1$ by considering a second vectorisation onto our super-operators, which act onto the doubly vectorised states denoted $|\cdot )$. We then have
\begin{align}
    |e^{(\frac{\Gamma_2}{2} - i\omega_2 + L)t_1 }  (a_2^\dagger)^T_L e^{Lt_1}) &= e^{L^T t_1}\otimes e^{(\frac{\Gamma_2}{2} - i\omega_2 + L)t_1 } |(a_2^\dagger)^T_L) \nonumber\\ & := e^{W t_1} |(a_2^\dagger)^T_L),
\end{align}
with $W = L^T \otimes \mathbb{1}_s + \mathbb{1}_s \otimes (\frac{\Gamma_2}{2} - i\omega_2 - L)$, and $\mathbb{1}_s$ is the identity superoperator. We can now perform the integral with respect to $t_1$ as
\begin{align}
    \int_0^{t_2}dt_1 & e^{(\frac{\Gamma_2}{2} - i\omega_2 + L)t_1}  (a_2^\dagger)^T_L e^{Lt_1} =  \int_0^{t_2}dt_1 e^{W t_1}|(a_2^\dagger)^T_L) \nonumber \\&
    = \frac{1}{W} (e^{Wt_2} - \mathbb{1}_s) |(a_2^\dagger)^T_L) \\ &
    = e^{(\frac{\Gamma_2}{2} - i\omega_2 + L)t_1 }  A  e^{Lt_1} - A\nonumber
\end{align}
where we have defined $A = W^{-1} (a_2^\dagger)^T_L $, which is a super-operator. Note that numerical determination of $A$ may become intractable due to the large matrix dimension of $W$; an issue we may circumvent by noting that $A$ is the solution to a Sylvester equation \cite{Bhatia1997Jan} of the form
\begin{equation}
    (a_2^\dagger)^T_L = W[A] = (\Gamma_2 / 2 - i\omega_2 - L)A + AL,
\end{equation}
which for our purposes is more amenable to numerical determination. We thus have
\begin{align}
    G_2(\tau) &= e^{-\Gamma_2 \tau} \int_0^\tau dt_2 A e^{(\Gamma_2 + L)t_2} - e^{(\frac{\Gamma_2}{2} + i\omega_2 + L)t_2}  A  \nonumber \\&
    = A\frac{1}{L + \Gamma_2}(e^{L\tau} -  e^{-\Gamma_2 \tau} \mathbb{1}) \nonumber \\ & \qquad 
    - \frac{1}{L + \frac{\Gamma_2}{2} + i\omega_2} (e^{(L - \frac{\Gamma_2}{2} + i\omega_2)\tau} - e^{-\Gamma_2 \tau} \mathbb{1}) A.
\end{align}
Thus, finally, we have
\begin{align}
    I_2(\tau) = -2 \epsilon^2 \textrm{Re}\bigg[I_{2a}(\tau) + I_{2b}(\tau) \bigg],
\end{align}
with,
\begin{align}
    I_{2a}(\tau) = \llangle a_2^\dagger | \frac{1}{L + \Gamma_2} &A  |\rho^{1, 0}_{1, 0}(\tau)\rrangle \nonumber \\& - e^{-\Gamma_2 \tau} \llangle a_2^\dagger |  \frac{1}{L + \Gamma_2} A |\rho^{1, 0}_{1, 0}(0)\rrangle 
\end{align}
\begin{align}
    I_{2b}(\tau) = e^{(i\omega_2 - \frac{\Gamma_2}{2})\tau}& \llangle a_2^\dagger | \frac{1}{L +\frac{\Gamma_2}{2} + i\omega_2}  |\chi(\tau)\rrangle \nonumber \\& - e^{-\Gamma_2 \tau} \llangle a_2^\dagger |  \frac{1}{L + \Gamma_2} A |\chi(0)\rrangle 
\end{align}
where we have defined $\chi(\tau) = e^{L \tau} A |\rho^{1, 0}_{1, 0}(0)\rrangle $. We can therefore evaluate $I_2(\tau)$ by computing the action of one super-duper-operator reciprocal, the action of two
super-operator reciprocals, and the dynamics of a modified reduced density matrix. Whilst this remains numerically challenging, we find that this approach is significantly more feasible than obtaining long-time dynamics with the integral approach. A Python implementation of this can be found at \cite{github}.

\section{Conventional quantum detailed balance measure}

\begin{figure}
    \includegraphics[width=0.98\linewidth]{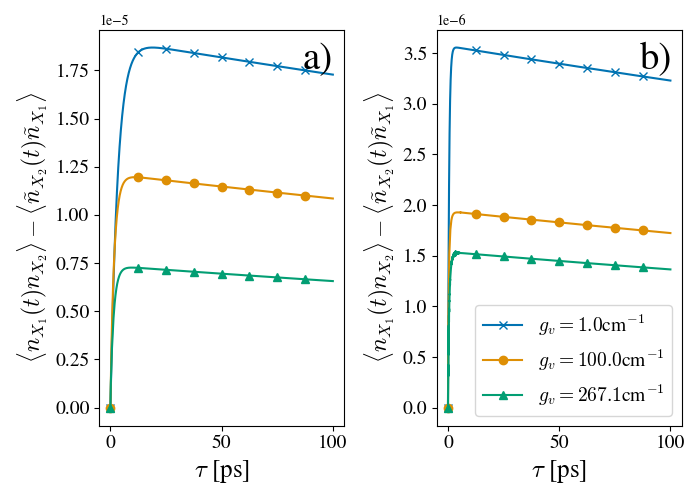}
    \caption{Conventional measure of QDB via Eq. \eqref{eq:qdb_def_SM} for a) dimer 1 and b) dimer 2. Here we show $A = n_{X_1} = |X_1\rangle \langle X_1|$ and $B = n_{X_2} = |X_2\rangle \langle X_2|$.}
    \label{fig:qdb_pops}
\end{figure}

In this section we provide some additional numerical results regarding the direct measure of quantum detailed balance in terms of observables and their time reversal, Eq. \eqref{eq:qdb_def} of the main text, which is repeated here for convenience:
 \begin{equation}\label{eq:qdb_def_SM}
    \langle A(t)B(0)\rangle \stackrel{QDB}{=} \langle \tilde{B}^\dagger(t) \tilde{A}^\dagger(0)\rangle \quad \forall \,\, A, B \in \mathcal{B(H)}.
\end{equation}
We note that as QBD implies this equality is satisfied for all operators $A, B$, there may be cases where chosen operators satisfy QDB, and others do not. In such cases the system as a whole violates QDB, but the chosen observables may obey this symmetry. Thus, the behaviour of QDB violation for differing observables may provide insight on the mechanisms of transport in a quantum system, as violation of QDB implies the formation of a current in the NESS. 

Firstly, in Fig. \ref{fig:qdb_pops} we show the case where the chosen observables are simply exciton populations, $A = n_{X_1} = |X_1\rangle \langle X_1|$ and $B = n_{X_2} = |X_2\rangle \langle X_2|$, and hence no coherences are probed by the QDB measure. We see the opposite tendency of QDB violation to that observed in the text, where increasing $g_v$ actually decreases the violation of QDB with this measure. This can be understood as the excitonic populations in this case evolve in a more reversible manner for higher $g_v$, as the Hamiltonian part of the open system dynamics contributes more to evolution. We see the measure of the main text increase with $g_v$, then, as it probes coherences directly, and $g_v$ opens a new coherent transfer pathway via the coherent transport mechanism explained in the main text.

\section{Choi-Jamio\l{}kowski isomorphism}\label{sec:choi}

The Choi-Jamio\l{}kowski isomorphism relates a linear map $M: \mathcal{H}_a \to \mathcal{H}_b$, where $\mathcal{H}_a$ and  $\mathcal{H}_b$ are not necessarily identical Hilbert spaces (but for our purposes will be), to a linear operator $O_{M} \in \mathcal{H}_{ab} := \mathcal{H}_a \otimes \mathcal{H}_b$. The space of linear operators on the combined space is given by $\mathcal{B}(\mathcal{H}_{ab})$, and is itself a Hilbert space with an inner product $\langle O_1 | O_2 \rangle = \Tr [O_1^\dagger O_2]$. We begin by writing the general form of the action of a map $M$ on a density operator $\rho \in \mathcal{H}_a$
\begin{align}\label{eq:map}
    M[\rho] = M\left[\sum_{ij} \rho_{ij} |i\rangle \langle j|\right] =  \sum_{ij} \rho_{ij} M\left[|i\rangle \langle j|\right] := \sum_{ij} \rho_{ij} C^{(M)}_{ij} .
\end{align}
Here $C^{(M)} \in \mathcal{H}_{ab}$, such that each entry $C^{(M)}_{ij} \in \mathcal{H}_b$. We can then note that the far right hand side of the above expression can itself be related to a trace operation, via $\Tr[AB] = \sum_{j}\langle j| A B |j \rangle = \sum_{ij}\langle j| A |i\rangle \langle i| B |j \rangle = \sum_{ij}A_{ij} B_{ji}$. Then, 
\begin{align}
    M[\rho] = \Tr_a[ (\rho^T \otimes \mathbb{1}_b) C^{(M)} ],
\end{align}
where $\mathbb{1}_b$ is the identity on $\mathcal{H}_b$. Note again that here $\rho \in \mathcal{H}_a$ and $C^{(M)} \in \mathcal{H}_{ab}$, such that $M[\rho] \in \mathcal{H}_b$, as required. $C^{(M)}$ is labelled the Choi state, as if, for example, the map $M$ is a completely positive trace preserving map, then $C^{(M)}$ is a valid density operator. The Choi state is often written in the form
\begin{align}
C^{(M)} = (\mathcal{I}_a \otimes M)[|\phi\rangle \langle \phi|],
\end{align}
with $|\phi\rangle = \sum_i |i\rangle |i\rangle \in \mathcal{H}_a \otimes \mathcal{H}_a$, and $\mathcal{I}_a: \mathcal{H}_a \to \mathcal{H}_a$ the identity map on $\mathcal{H}_a$ such that
\begin{align}
    C^{(M)} = \sum_{ij} | i \rangle \langle j | \otimes M[| i \rangle \langle j |],
\end{align}
from which we can recover Eq. \eqref{eq:map} via
\begin{align}
    \sum_{ij} \rho_{ij} \langle i| C^{(M)} |j\rangle &= \sum_{ij} \rho_{ij}  \langle i | \sum_{kl} | k \rangle \langle l | \otimes M[| k \rangle \langle l |] | j\rangle \nonumber \\ &
    = \sum_{ij}\rho_{ij} M[|i\rangle \langle j|] = M[\rho].
\end{align}
The Choi-Jamio\l{}kowski isomorphism is thus a statement that the information contained in the map $M$ similarly contained in an operator $C^{(M)}$ on an extended Hilbert space.

\subsection{Time-Reversal}\label{sm:time_rev}

The notion of QDB in Eq. \eqref{eq:qdb_def} relies on the anti-linear (anti-unitary and linear) time reversal operator $T$. The anti-linearity implies the following properties:
\begin{align}
    &T(A+B) = TA + TB\\&
    T(\alpha A) = \alpha^* TA,
\end{align}
where $A, B$ are operators and $\alpha$ is a complex scalar.
For unitary quantum dynamics, the time reversal operator $T = K$, where $K$ is the complex conjugation operator, more generally, $T = UK$, where $U$ is some unitary. In general all anti-unitary operators may be decomposed as a product of unitary and anti-unitary operators in this manner.

We refer to the time-reversed operators as $\tilde{A} = TAT^{-1}$. It can then be seen that the time reversal of any quantum process is equal to its forward process via cyclic permittivity of the trace $\Tr [ \widetilde{\cdots}] = \Tr[ T \cdots T^{-1}] = \Tr[\cdots]$.

The above considerations lead to some key features of time reversal in quantum theory: 1) The preservation of transition probabilities, $|\langle \psi | \phi \rangle| = |\langle T\psi| T\phi\rangle|$. 2) Observables have some parity under time reversal, $\tilde{O} = TOT^{-1} = \sum_{ij} O_{ij} T|i\rangle \langle j| T^{-1} = \sum_{ij} O_{ij} T(\sum_{k} c_{ik} |i\rangle)(\sum_{l} c_{jl}^* \langle j|)T^{-1} =  \sum_{ij} O_{ij} (\sum_{k} c^*_{ik} |i\rangle)(\sum_{l} c_{jl} \langle j|) =  \sum_{ij} \sum_{kl} O_{ij} c^*_{ik} c_{jl} |i\rangle\langle j| = \sum_{ij} \sum_{kl} O_{ij} e^{i(\phi_{ik} - \phi_{kl})}|c_{ik}| |c_{jl}| |i\rangle\langle j|$ compared to the forward time version $O = \sum_{ij}  \sum_{lk} O_{ij}c_{ik} c^*_{jl}  |i\rangle\langle j| = \sum_{ij} \sum_{kl} O_{ij} e^{-i(\phi_{ik} - \phi_{kl})} |c_{ik}| |c_{jl}| |i\rangle\langle j|$. 

We note the anti-unitary complex conjugation operation is \textit{basis-dependent}. We implement the conjugation in the site-basis.

\end{widetext}

\end{document}